\documentclass[useAMS,usenatbib]{mn2e}
\usepackage{graphicx}
\usepackage[T1]{fontenc}
\usepackage{aecompl}
\usepackage{times}
\usepackage{amsmath,amssymb}
\usepackage{rotating}
\usepackage{multirow}
\title[The physical properties of \textit{Fermi} TeV BL Lacs jets]{The physical properties of \textit{Fermi} TeV BL Lac objects jets}
\author[N. Ding, X. Zhang, D. R. Xiong, H. J. Zhang]{
N. Ding$^{1}$, X. Zhang$^{1}$\thanks{E-mail:ynzx@yeah.net}, D. R. Xiong$^{2}$, H. J. Zhang$^{1}$\\
$^{1}$Department of physics, Yunnan Normal University, Kunming 650500, China\\
$^{2}$Yunnan observatories, Chinese Academy of Sciences, Kunming 650011, China\\}
\begin{document}

\maketitle

\label{firstpage}

\begin{abstract}
We investigate the physical properties of \textit{Fermi} TeV BL Lac objects jets by modeling the quasi-simultaneous spectral energy distribution of 29 \textit{Fermi} TeV BL Lacs in the frame of a one-zone leptonic synchrotron self-Compton model. Our main results are the following: (i) There is a negative correlation between $B$ and $\delta$ in our sample, which suggests that $B$ and $\delta$ are dependent on each other mainly in Thomson regime. (ii) There are negative correlations between $\nu_{\text{sy}}$ and $r$, the $\nu_{\text{IC}}$ and $r$, which is a signature of the energy-dependence statistical acceleration or the stochastic acceleration. There is a significant correlation between $r$ and $s$, which suggests that the curvature of the electron energy distribution is attributed to the energy-dependence statistical acceleration mechanism. (iii) By assuming one proton per relativistic electron, we estimate the jet power and radiative power. A size relation $P_{\text{e}} \sim P_{\text{p}} > P_{\text{r}} \gtrsim P_{\text{B}}$ is found in our sample. The $P_{\text{e}}>P_{\text{B}}$ suggests that the jets are particle dominated, and the $P_{\text{e}}\sim P_{\text{p}}$ means that the mean energy of relativistic electrons approaches $m_{\text{p}}/m_{\text{e}}$. There are not significant correlations between $P_{\text{jet}}$ and black hole mass in high or low state with a sub-sample of 18 sources, which suggests that the jet power weakly depends on the black hole mass. (iv) There is a correlation between the changes in the flux density at 1 TeV and the changes in the $\gamma_{\text{peak}}$, which suggests the change/evolution of electron energy distribution may be mainly responsible for the flux variation.
\end{abstract}


\section{Introduction}
Blazars are the most extreme active galactic nuclei (AGNs) pointing their jets in the direction of the observer (Urry \& Padovani 1995). They have high luminosity, large amplitude and rapid variability, high and variable polarization, radio core dominance, and apparent super-luminal speeds (Urry \& Padovani 1995; Massaro et al. 2016). Generally, Blazars are divided into subcategories of BL Lacs objects (BL Lacs), characterized by almost completely lacking of emission lines or only showing weak emission lines (EW $\leq$ 5 $\mathring{\text{A}}$), and highly polarized quasars or flat spectral radio quasars (FSRQs), showing broad strong emission lines (Falomo et al. 2014; Massaro et al. 2014). The broadband spectral energy distributions (SEDs) of blazars are double peaked. The bump at the IR-optical-UV band is explained with the synchrotron emission of relativistic electrons, and the bump at the GeV-TeV gamma-ray band is due to the inverse Compton (IC) scattering (e.g., Dermer et al. 1995; Dermer et al. 2002; B\"{o}ttcher 2007). The seed photons for IC process could be from the local synchrotron radiation on the same relativistic electrons (i.e. synchrotron self-Compton (SSC); e.g., Tavecchio et al. 1998), or from the external photon fields (EC; e.g., Dermer et al. 2009), such as those from accretion disk (e.g., Dermer \& Schlickeiser 1993) and broad-line region (e.g., Sikora et al. 1994). The hadronic model is an alternative explanation for the high energy emissions from blazars (e.g., Dermer et al. 2012).

The modeling of SED with a given radiation mechanism allow us to investigate the intrinsic physical properties of emitting region and the physical conditions of jet (e.g., Ghisellini \& Tavecchio 2008; Celotti \& Ghisellini 2008; Ghisellini et al. 2009; Ghisellini et al. 2010; Ghisellini et al. 2011; Zhang et al. 2012; Yan et al. 2014). Celotti \& Ghisellini (2008) estimated the powers of blazars jets based on EGRET observations and they found that the typical jet should comprise an energetically dominant proton component and only a small fraction of the jet power is radiated if there is one proton per relativistic electron. In addition, in their work, the TeV BL Lacs shows some special jet properties, such as $P_{e} \sim P_{p}$ and relatively high radiation efficiency. Ghisellini et al. (2009, 2010, 2011) mainly concerned the relation between the jet power and the accretion disk luminosity in \textit{Fermi} blazars and they found that there is a positive correlation between the jet power and the accretion disk luminosity for \textit{Fermi} broad-line blazars.

The $\gamma$-ray extragalactic sky at high (> 100 MeV) and very high (> 100 GeV) energies is dominated by blazars. About 50 blazars have been detected in the TeV gamma-ray band\footnote[1]{http://tevcat.uchicago.edu/.}, and most of them are the BL Lacs. The SEDs of TeV BL Lacs suffer less contamination of the emission from the accretion disk and EC process, so it can be explained well by the one-zone SSC model (Paggi et al. 2009a; Dermer et al. 2015). Since the launch of the \textit{Fermi} satellite, we have entered in a new era of blazar research (Abdo et al. 2009; Abdo et al. 2010a). The abundant data observed by \textit{Fermi}/LAT in the MeV-GeV band, together with the multi-wavelength campaigns at the radio, optical, X-ray bands and the ground-based observations at the TeV gamma-ray band, now provide an excellent opportunity to study the TeV blazars (e.g., Massaro et al. 2011a, Giommi et al. 2012; Massaro et al. 2013).

In this paper, we have collected the quasi-simultaneous broadband SEDs of 29 \textit{Fermi} TeV BL Lacs from the literatures. We used the one-zone leptonic synchrotron self-Compton model with the log-parabolic electron energy distribution to fit SEDs. And we used the Markov Chain Monte Carlo sampling method instead of the "eyeball" fitting to obtain the best-fit model parameters. Then, based on the model parameters, we systematically investigated the physical properties of \textit{Fermi} TeV BL Lacs jets through statistical analysis. This paper is organized as follows: In Sect.2, we present the sample, the model and the fitting strategy are presented in Sect.3. Then, results and discussions are showed in Sect.4. Finally, we end with a conclusion of the findings in Sect.5. The cosmological parameters $H_{0}=70~$Km~s$^{-1}$Mpc$^{-1}$, $\Omega_{m}=0.3$, and $\Omega_{\Lambda}=0.7$ are adopted in this work.
\section{THE SAMPLE}
Up to now, 55 BL Lacs have been detected in the TeV or very high energy regime$^1$. 46 of them have been detected at MeV-GeV energies by \textit{Fermi}/LAT. In order to study the physical properties of \textit{Fermi} TeV BL Lacs jets, we have collected the quasi-simultaneous broadband SEDs of 29 \textit{Fermi} TeV BL Lacs from the literatures. For each source, the time of quasi-simultaneous observation, the observation telescopes, and the references are listed in Table 1. Our sample accounts for about 65\% of the total number of \textit{Fermi} TeV BL Lacs. There are 12 sources which have two broadband SEDs. The two SEDs for 12 sources are defined as a high or low state with the observed or extrapolated flux density at 1 TeV. The observed broadband SEDs are shown in Figure 1.

\begin{figure*}
\centering
\includegraphics{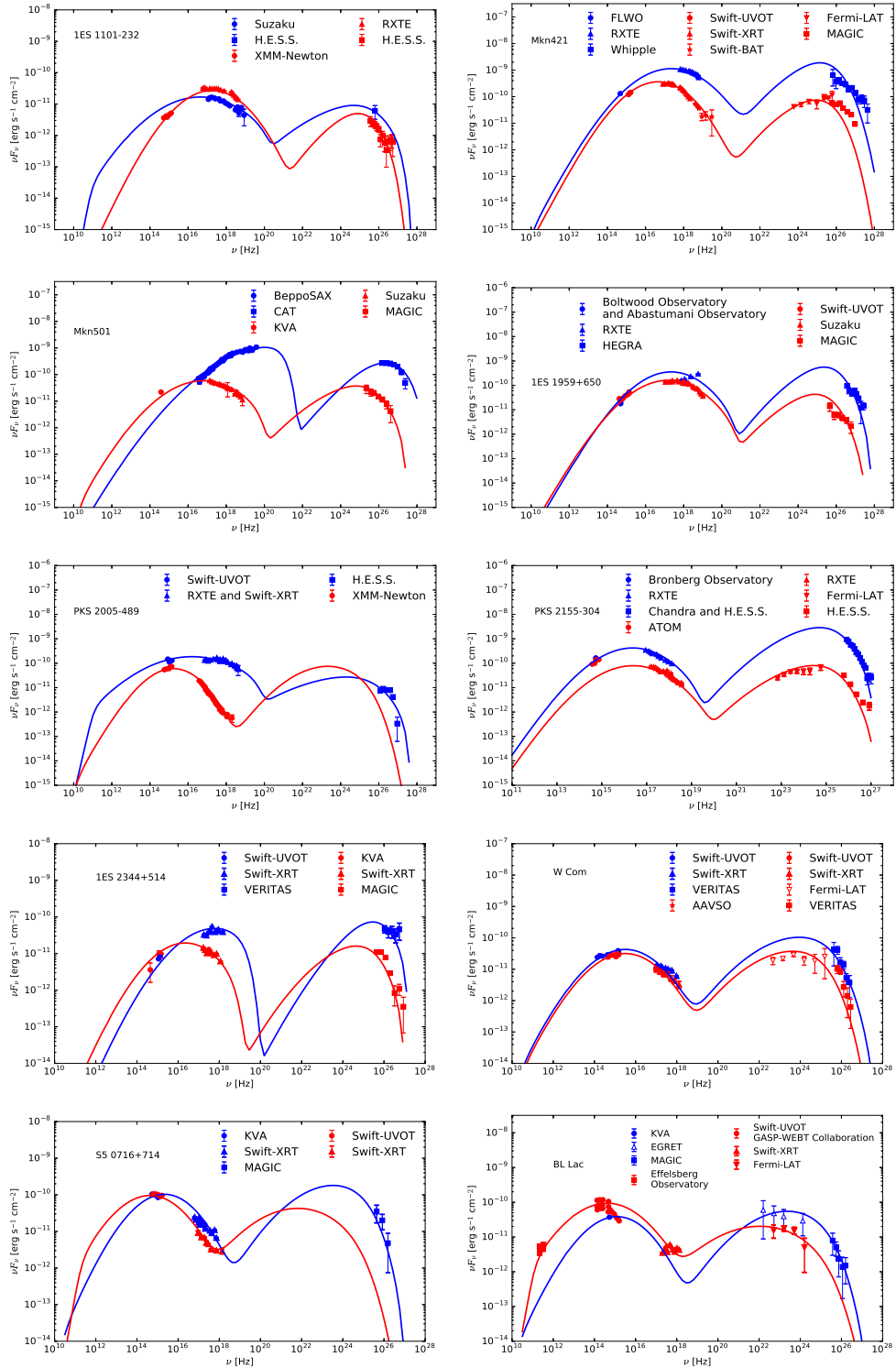}
\caption{The observed SEDs (scattered data points) with our best model fittings (lines) of 29 \textit{Fermi} TeV BL Lacs. The data of high and low states are marked with blue and red symbols, respectively. If only one SED is obtained, the data are shown with black symbols. The open symbols are for the data obtained not in the time of quasi-simultaneous observation, the detailed information about the data sets are listed in Table 1. For Mkn180, 3C 66A, PKS 1424+240 and 1ES 1215+303, the red dashed lines are calculated by the on-line SSC calculator (see text in detail).}
\end{figure*}
\begin{figure*}
\addtocounter{figure}{-1}
\centering
\includegraphics{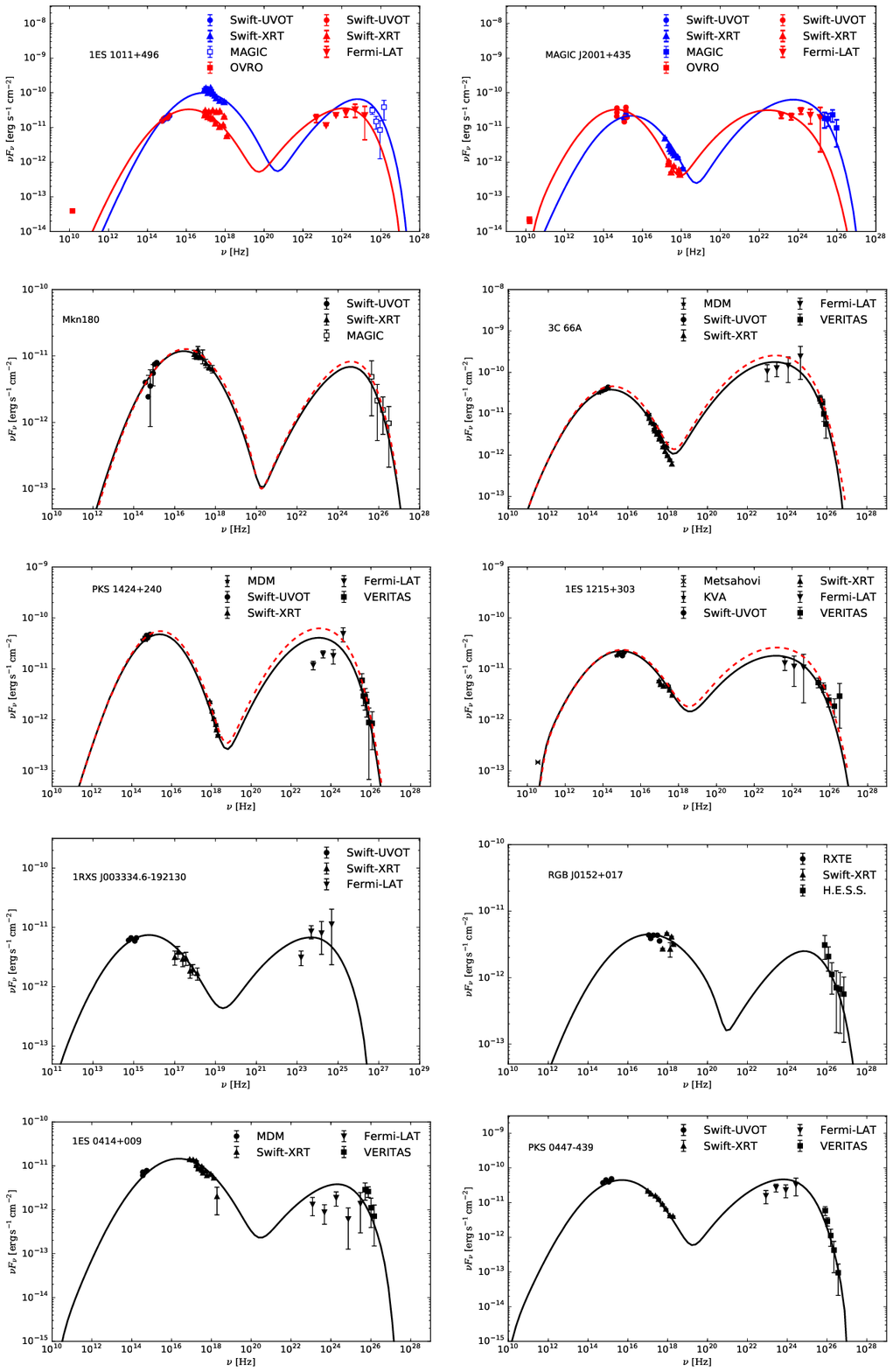}
\caption{Continued}
\end{figure*}
\begin{figure*}
\addtocounter{figure}{-1}
\centering
\includegraphics{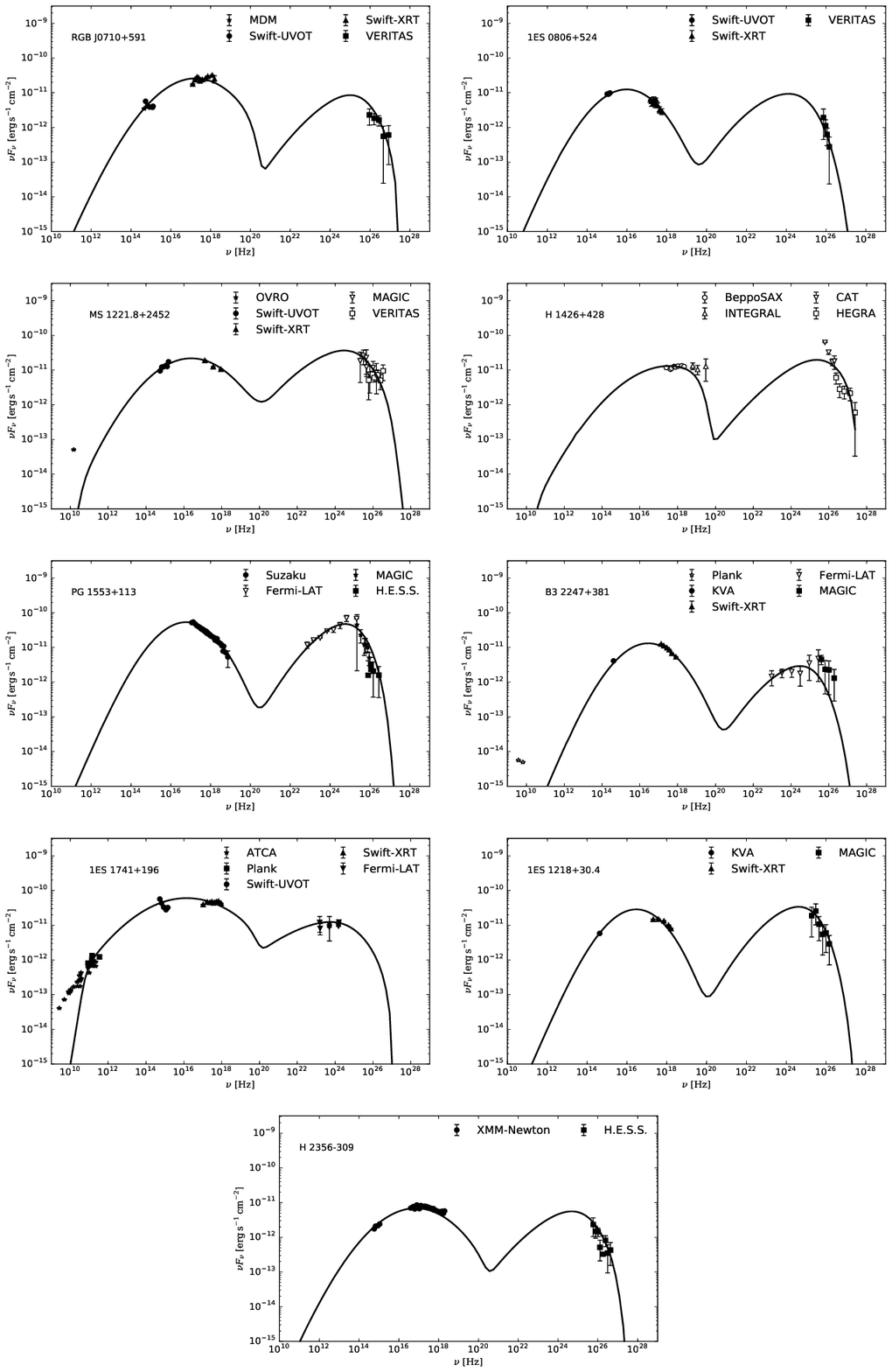}
\caption{Continued}
\end{figure*}
\section{MODELING FITTING PROCEDURE}
\subsection{Model}
We adopt a one-zone leptonic SSC model. And we assume the electron energy distribution is a log-parabolic spectra (Massaro et al. 2004a; Tramacere et al. 2011),
\begin{equation}
N(\gamma ) = N_0 (\frac{\gamma }{\gamma _0 })^{ - s -
r \mbox{log}(\frac{\gamma }{\gamma _0 })} = NK(\frac{\gamma }{\gamma _0 })^{
- s - r\mbox{log}(\frac{\gamma }{\gamma _0 })}.
\end{equation}
$K$ is the normalization coefficient, 
\begin{equation}
\int_{\gamma _{\min } }^{\gamma _{\max } } {K(\frac{\gamma }{\gamma _0 })^{- s - r \mbox{log}(\frac{\gamma }{\gamma _0 })}} \mbox{ = }1.
\end{equation}
$N$ is the number of emitting particles per unit volume expressed in $1/$cm$^{3}$. $\gamma_{0}$ is the reference energy. $r$ is the spectral curvature. $s$ is the spectral index at the reference energy $\gamma_{0}$. $\gamma_{\min}$ and $\gamma_{\max}$ are the minimum and maximum energies of electrons. The log-parabolic electron energy distribution can be obtained from a Fokker-Planck equation as first shown by Kardashev(1962). The log-parabolic electron energy distribution provides more accurate fits than broken power-laws in the optical X-ray bands, with residuals uniformly low throughout a wide energy range (Paggi et al. 2009a). Recently, the log-parabolic electron energy distribution has also been applied to describe the spectral distribution and evolution of some gamma-ray bursts (Massaro \& Grindlay 2011).

The radiation region is taken as a homogeneous sphere with radius $R$ and filled with the uniform magnetic field $B$. Due to the beaming effect, the observed radiation is strongly boosted by relativistic Doppler factor $\delta$. The synchrotron emission coefficient is calculated with
\begin{equation}
j_{\text{syn}} (\nu ) = \frac{1}{4\pi }\int_{\gamma _{\min } }^{\gamma _{\min
} } {N(\gamma )P(\nu ,\gamma )d} \gamma,
\end{equation}
and the synchrotron absorption coefficient is calculated with
\begin{equation}
\alpha _{\text{syn}} (\nu ) = - \frac{1}{8\pi \nu ^2m_e }\int_{\gamma _{\min }
}^{\gamma _{\max } } {d\gamma P(\nu ,\gamma )\gamma ^2\frac{\partial
}{\partial \gamma }} [\frac{N(\gamma )}{\gamma ^2}].
\end{equation}
Where $P(\nu,\gamma)$ is the single electron synchrotron emission averaged over an isotropic distribution of pitch angles (see, e.g., Ghisellini et al. 1988), $m_{\text{e}}$ is electron mass. According to the radiative transfer equation of spherical geometry, we can calculate the synchrotron radiation intensity,
\begin{equation}
I_{\text{syn}} (\nu ) = \frac{j_{\text{syn}} (\nu )}{\alpha _{\text{syn}} (\nu )}[1 -
\frac{2}{\tau(\nu) ^2}(1 - \tau e^{ - \tau(\nu) } - e^{ - \tau(\nu) })],
\end{equation}
where $\tau(\nu) = 2R\alpha _{\text{syn}} (\nu)$.
The IC emission coefficient is calculated with 
\begin{equation}
j_{\text{IC}} (\epsilon ) = \frac{h\epsilon }{4\pi }\int d \epsilon _0
n(\epsilon _0 )\int {\gamma N(\gamma )}C(\epsilon ,\gamma ,\epsilon _0 ),
\end{equation}
where $\epsilon$ is scattered photon  energy and $\epsilon _0$ is soft photon energy (in unit of $m_{e}c^{2}$). $C(\varepsilon ,\gamma ,\epsilon _0)$ is the Compton kernel of Jones (1968), $n(\epsilon _0)$ is the number density of synchrotron soft photons per energy interval. More detailed calculation process see Kataoka et al. (1999). Due to the medium is transparent for IC radiation, hence $I_{\text{IC}}(\nu)=j_{\text{IC}}(\nu)R$.

Assuming that $I_{\text{syn, IC}}$ is an isotropic radiation field, the total observed flow density is given by 
\begin{equation}
F_{\text{obs}}(\nu_{\text{obs}}) = \frac{\pi R^2\delta ^3(1 + z)}{d_L ^2}(I_{\text{syn}} (\nu ) + I_{\text{IC}} (\nu )),
\end{equation}
where $d_{L}$ is the luminosity distance, $z$ is the redshift, and $\nu_{\text{obs}} = \nu \delta /(1+z)$. In addition, the high-energy gamma-ray photons can be absorbed by extragalactic background light (EBL), yielding electron-positron pairs. It makes the observed spectrum in the very high energy (VHE) band must be steeper than the intrinsic one. Therefore, according to the average EBL model in Dwek \& Krennrich (2005), we calculate the absorption in the GeV-TeV band.
\begin{figure*}
\centering
\includegraphics[width=10cm,height=6cm]{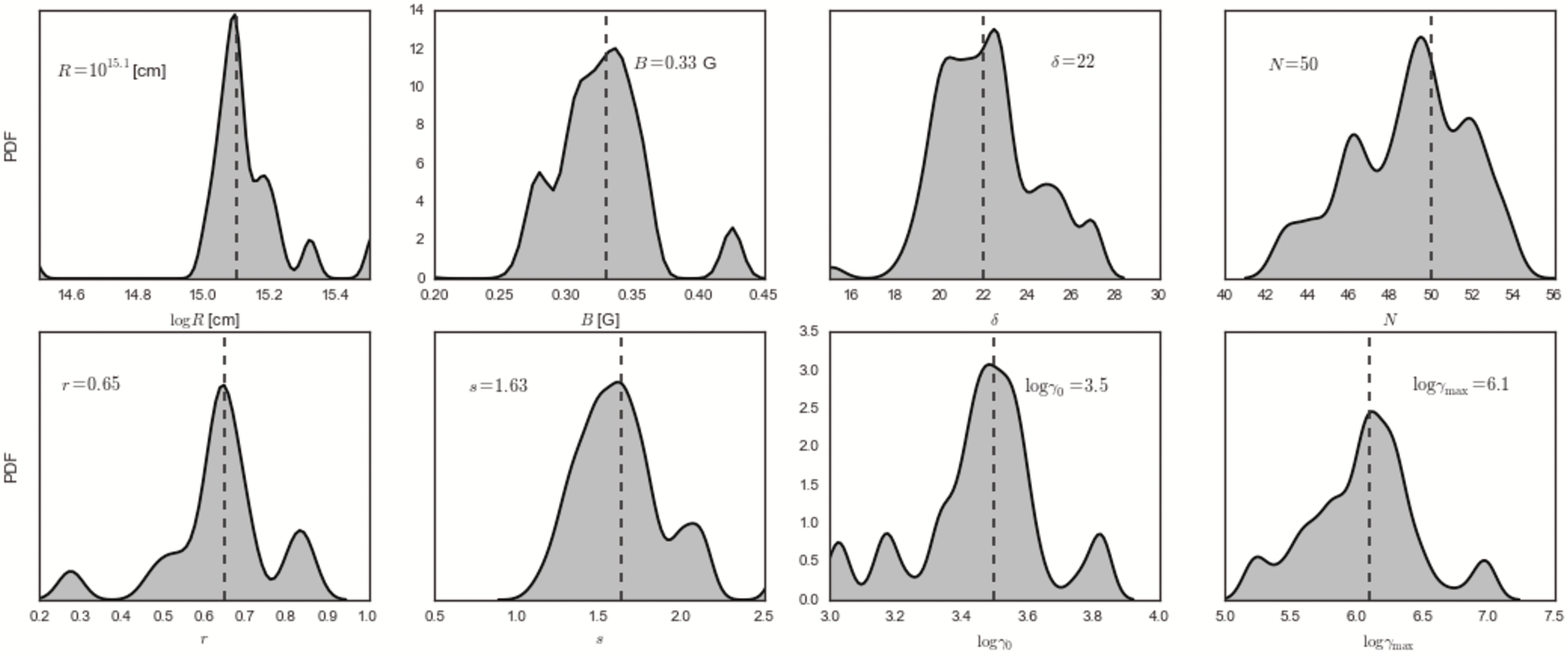}
\caption{The posterior distributions of model parameters for Mrk180 obtained through the MCMC sampling. The dotted lines are the median lines. we use the posterior medians as the best-fit values and estimate uncertainties based on the 16th and 84th percentiles.}
\end{figure*}
\subsection{Fitting strategy}
In this model, there are nine free parameters. Six of them specify the electron energy distribution ($N,\gamma_{\min},\gamma_{\max},\gamma_{0},r,s$), and other three ones describe the properties of the radiation region ($R,B,\delta$). For a given source, we calculate its non-thermal flux to fit the observed multi-wavelength data. We derive the best-fit and uncertainty distributions of model parameters through Markov Chain Monte Carlo (MCMC) sampling of their likelihood distributions. When measurements and uncertainties in the observed multi-wavelength data are assumed to be correct, Gaussian, and independent. The reduced log-likelihood of observed data given the model flux $S(\vec{p},\nu)$, for a parameter vector $\vec{p}$, is
\begin{equation}
\ln\mathcal{L} \propto  \sum^N_{i=1} \frac{(S(\vec{p},\nu_{i}) - F_i)^2}{\sigma^2_i},
\end{equation}
where $F_i,\sigma_i$ are the flux measurement and uncertainty at a frequency $\nu_{i}$ over $N$ spectral measurements. Combining the prior distributions of model parameters and the reduced log-likelihood function (8), we can use the code \textbf{emcee} (Foreman-Mackey et al. 2012), which implements the affine-invariant ensemble sampler of Goodman \& Weare (2010), to perform the MCMC algorithm. More details about this application can be found in Zabalza (2015). In concrete calculation, we use the uniform distribution as the prior distribution for all model parameters. The MCMC algorithm requires the initial input values of model parameters, at beginning we do a preliminary modeling to the SED for each object to guess the starting values. In addition, when the observational data are not given the uncertainty, we use the observational flux of 1\% as the error in radio band, and use the observational flux of 2\% as the error in optical to X-ray band, for $\gamma$-ray band, we estimate the error with the average relative error of the data points whose errors are available (Zhang et al. 2012; Aleksi\'c et al. 2014a). 

The minimum energy of electrons $\gamma_{\min}$ is historically poorly constrained by SED modeling, especially for low-power BL Lacs (Celotti \& Ghisellini 2008). The same as the previous studies (Rani et al. 2011; Yan et al. 2014; Kang et al. 2016), we use a fixed $\gamma_{\min}$ in our fitting. Due to the synchrotron self-absorption, the radio emission cannot be used to constrain the value of $\gamma_{\min}$. However, the observed spectral index between the X-ray and the $\gamma$-ray and the GeV data at low energies could place constraints on $\gamma_{\min}$ in some degree (Tavecchio et al. 2000; Yan et al. 2014). According to the average observed spectral index between the X-ray and the $\gamma$-ray of the high peaked frequency BL Lacs $\alpha_{\text{X}\gamma}$=1.02 (Fan et al. 2012) and the GeV data at low energies, we do a preliminary modeling to the SED for each object. Based on the preliminary modeling result, the $\gamma_{\min} = 10$ are adopted in our fitting. Figure 2 shows the posterior distributions of model parameters for Mrk180 obtained through the MCMC sampling. We use the posterior medians as the best-fit values and estimate uncertainties based on the 16th and 84th percentiles.
\section{RESULTS AND DISCUSSION}
Using above mentioned model and fitting strategy, we fit the 29 \textit{Fermi} TeV BL Lacs and obtain their best-fit model parameters (see Figure 1 and Table 2). Due to the synchrotron self-absorption, the one-zone SSC emission will fall in the radio band. It leads to the observed data in the radio band can not be well fitted. The one-zone SSC model here adopted is aimed to explain the bulk of the emission in compact radiation regions, but the radio flux mainly comes from a large-scale radiation regions outside the jet, so the radio data are not restricted by the one-zone SSC model, meanwhile, the bad fitting in the radio band does not affect the determination of model parameters (Ghisellini et al. 2009; Ghisellini et al. 2010).

Tramacere et al. provided an on-line SSC calculator\footnote[2]{http://www.isdc.unige.ch/sedtool/.} (Tramacere et al. 2007a, 2009). We use the same model parameters to calculate SSC emission by using this on-line SSC calculator for Mkn180, 3C 66A, PKS 1424+240 and 1ES 1215+303. The results are plotted as the red dashed lines in Figure 1. It can be found that the results are similar to ours, which proves that our SSC code is reliable. The minor differences in results are mainly caused by the different approximation methods for Bessel function. In our work, we use the approximate method proposed by Aharonian et al. (2010), the approximation error is less than 0.2\%.

\subsection{The distributions}
In Figure 3, we give the distributions of the spectral curvature $r$ and the spectral index $s$. It is found that the range of $r$ are from 0.30 to 1.21, the values of $s$ are clustered at 1.10 $\sim$ 2.30. The mean values of $r$ and $s$ are 0.79 and 1.85, respectively.

We show the distributions of $R$, $B$, and $\delta$ in Figure 4. The values of $R$ are in the range of (0.7 $\sim$ 40) $\times$ $10^{15}$ cm, which is consistent with Zhang et al (2012). The values of magnetic field in the emitting region $B$ are in the range of 0.1 $\sim$ 1 G, which is consistent with Ghisellini et al. (2011) and Zhang et al.(2012). The values of $\delta$ are in the range of 8 $\sim$ 38, which is similar to the observed results for $\gamma$-loud blazars (L\"{a}hteenmaki \& Valtaoja 2003; Savolainen et al. 2010; Lister et al. 2013; Lister et al. 2016). We also show the distributions of the peak frequency of synchrotron radiation $\nu_{\text{sy}}$ and the peak frequency of inverse Compton scattering $\nu_{\text{IC}}$ in Figure 4 (d). It is found that the $\nu_{\text{sy}}$ range from the infrared to X-ray band (from $10^{14}$ Hz to $10^{20}$ Hz). The values of $\nu_{\text{IC}}$ cover a range from $10^{20}$ Hz to $10^{27}$ Hz. 

Finke et al. (2013) used a sub-sample of the second LAT catalog (2LAC) to investigated the compton dominance and blazar sequence. In their work, they obtained the values of synchrotron peak frequency ($\nu^{\prime}_{\text{sy}}$) form Ackermann et al. (2011) (Ackermann et al. (2011) used a empirical relation between the spectral index and synchrotron peak frequency to estimate the value of synchrotron peak frequency), and obtained the values of IC peak frequency ($\nu^{\prime}_{\text{IC}}$) by using a 3rd-degree polynomial to fit the non-simultaneous SEDs. By cross-correlating their sample with ours, we get the values of $\nu^{\prime}_{\text{sy}}$ and $\nu^{\prime}_{\text{IC}}$ of 19 sources obtained by Finke et al. (2013), and we compare these values with ours (see Figure 5). It can be found the values of the $\nu^{\prime}_{\text{sy}}$ and $\nu^{\prime}_{\text{IC}}$ are only roughly similar to ours, the result shows that the estimations of the synchrotron peak frequency and IC peak frequency are greatly influenced by the SED data and the estimation method as mentioned by Finke et al. (2013). 

\subsection{Magnetic field vs Doppler factor}
We plot the $\delta$ as a function of the $B$ in Figure 6. We find that a negative correlation between $B$ and $\delta$ with the Pearson correlation coefficient $\rho=-0.47$ and the chance probability $P = 0.0021$ for our sample. In addition, we do a covariance analysis and a partial correlation analysis to exclude possible covariates (other model parameters) effects. The results of the analyses are listed in Table 4. From the results, we verify that the $B$-$\delta$ correlation is reliable. According to the SSC model, there are relationships between $B$ and $\delta$ for a given source, $B\delta \propto [\nu_{\text{sy}}^{2}/\nu_{\text{IC}}](1+z)$ in Thomson regime and $B/\delta \propto [\nu_{\text{sy}}/\nu_{\text{IC}}^{2}]/(1+z)$ in the Klein-Nishina (KN) regime (Tavecchio et al. 1998; Yan et al. 2014). Our result is in agreement with the SSC model prediction and suggests that $B$ and $\delta$ are dependent on each other mainly in Thomson regime. Yan et al. (2014) explored the correlation between $B$ and $\delta$. Their sample consists of 10 high-synchrotron peaked BL Lacs, 6 intermediate-synchrotron peaked BL Lacs and 6 low-synchrotron peaked BL Lacs. Because the synchrotron peaks of their sources distribute in a large range, the values of $[\nu_{\text{sy}}^{2}/\nu_{\text{IC}}]$ and the $[\nu_{\text{sy}}/\nu_{\text{IC}}^{2}]$ are relatively scattered. It causes they did not find any correlations between $B$ and $\delta$.

\begin{figure}
\centering
\includegraphics[width=12cm,height=9.5cm]{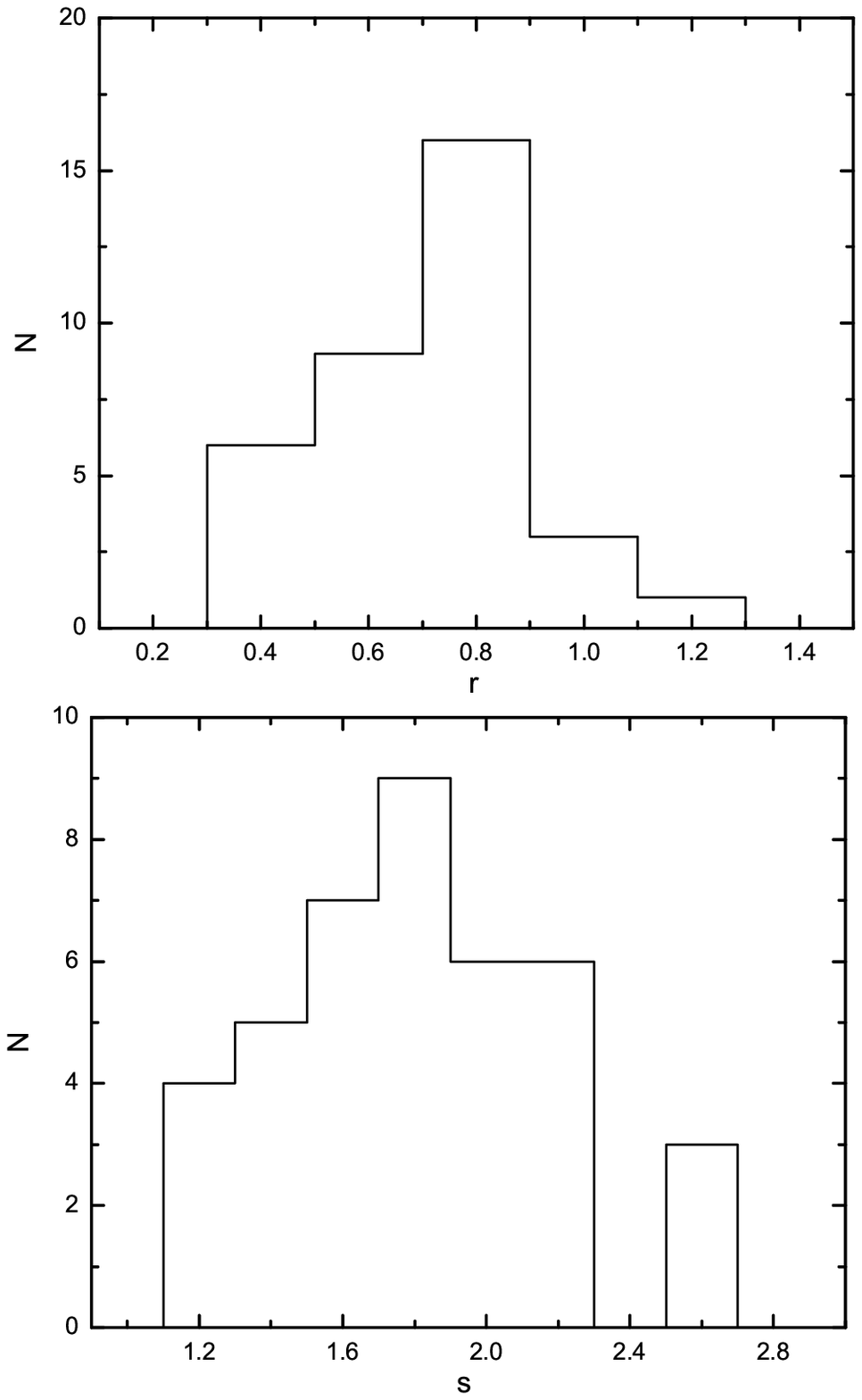}
\caption{Distributions of the spectral curvature $r$ and the spectral index $s$.}
\end{figure}
\begin{figure}
\centering
\includegraphics[width=12cm,height=9.5cm]{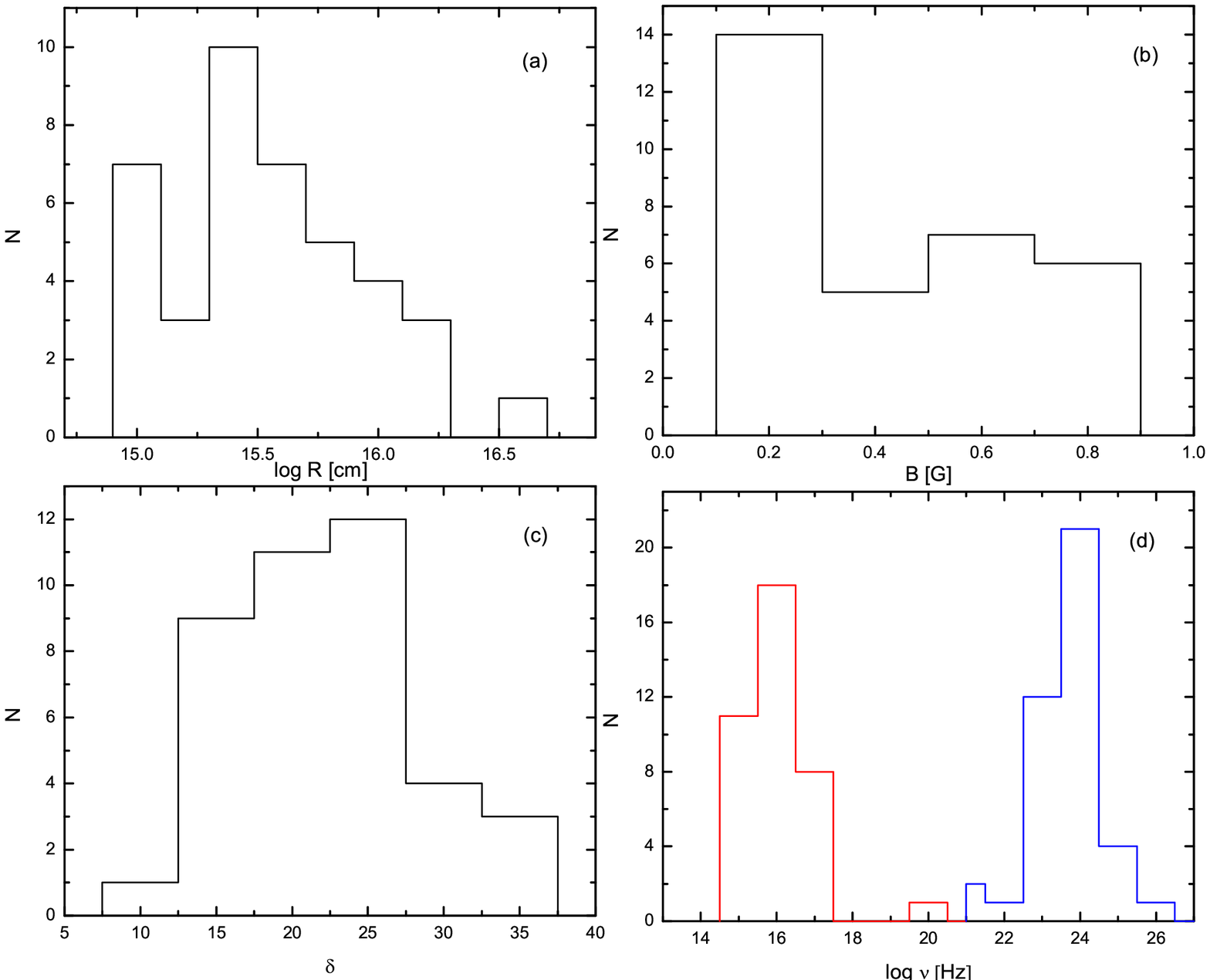}
\caption{Distributions of the radiation region radius $R$ (a), the magnetic field $B$ (b), the beaming factor $\delta$ (c). (d) Distribution of the synchrotron radiation peak frequency $\nu_{\text{sy}}$ (red solid line) and the inverse Compton scattering peak frequency $\nu_{\text{IC}}$ (blue solid line).}
\end{figure}
\begin{figure}
\centering
\includegraphics[width=12cm,height=9.5cm]{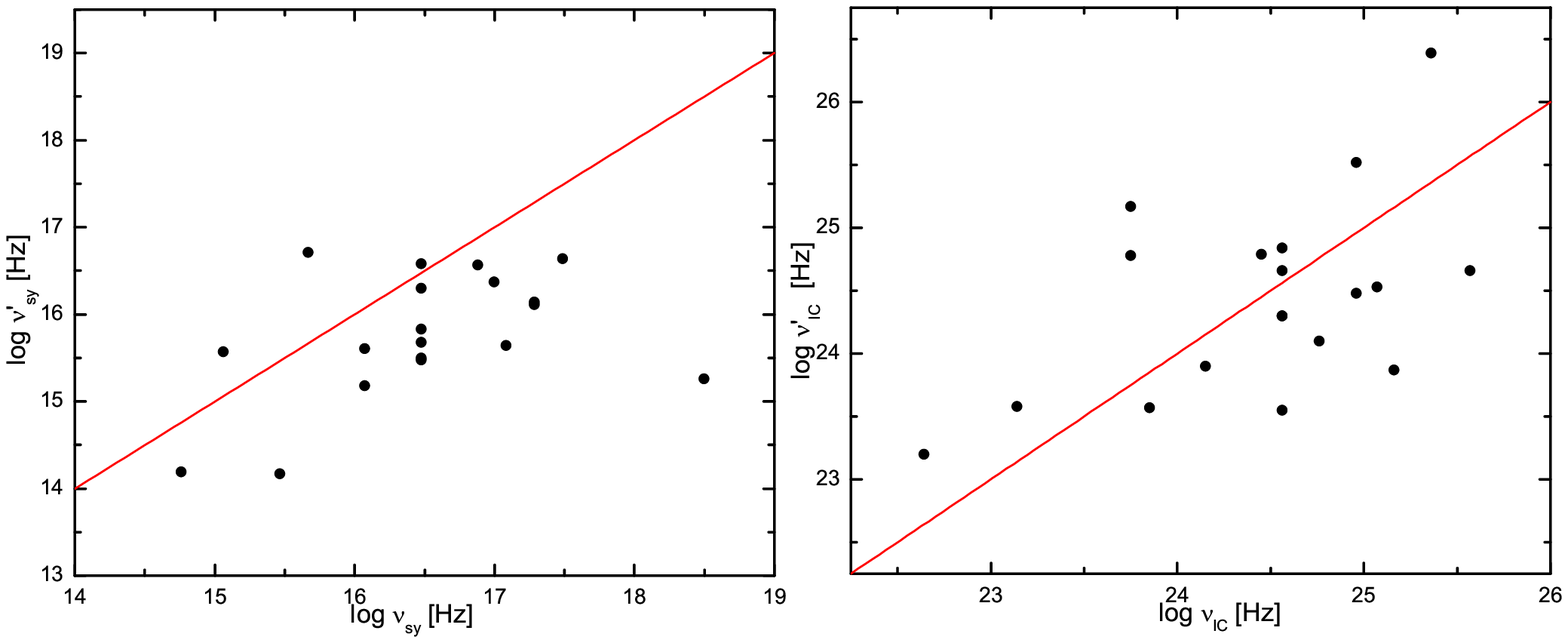}
\caption{The comparisons between $\nu_{\text{sy}}$, $\nu_{\text{IC}}$ obtained by our and $\nu^{\prime}_{\text{sy}}$, $\nu^{\prime}_{\text{IC}}$ obtained by Finke et al. (2013). The red lines are $y=x$. In here, we take the average value for those sources that have the high and low state in our sample.}
\end{figure}

\subsection{The particle acceleration and the radiative cooling}
In the framework of statistical acceleration, Massaro et al. (2004a, 2004b, 2006) showed that when the acceleration efficiency is inversely proportional to the acceleration particles's energy itself, the electron energy distribution is curved into a log-parabolic shape, and its curvature $r$ is related to the fractional acceleration gain $\epsilon$,
\begin{equation}
r \propto \frac{1}{{\log \epsilon}}.
\end{equation}
$r$ decreases when $\epsilon$ increases. Note that the $E_{\text{sy, IC}} = h \nu_{\text{sy, IC}}$ scales like $\epsilon$, where $h$ is Planck constant. Hence, according to this model, an inverse correlation between $\nu_{\text{sy, IC}}$ and $r$ is expected. 

In addition, a connection of log-parabolic electron energy distribution with acceleration can be also understood in the framework provided by the Fokker-Planck equation. In this case, the diffusion term acts on the electron spectral curvature as given by Melrose (1969) and Kardashev (1962),
\begin{equation}
r \propto \frac{1}{{Dt}},
\end{equation}
where $D$ is the diffusion term. This relation will lead to the inverse correlation between $\nu_{\text{sy ,IC}}$ and $r$ (see Tramacere et al. 2007a and Tramacere et al. 2009). So, the inverse correlation between $\nu_{\text{sy ,IC}}$ and $r$ also is a strongest signature of the stochastic acceleration process (Tramacere et al. 2011). The statistical and stochastic acceleration process imply that the curvature of radiation spectrum are not simply the result of radiative cooling of high energy electrons, responsible for the synchrotron and IC emission, but is essentially related to the acceleration mechanism (Massaro 2004a, Massaro et al. 2011b). The curvature basically arises from a loss of acceleration efficiency at high energies.

We plot the $\nu_{\text{sy, IC}}$ vs $r$ in Figure 7. We find a negative correlation between $\nu_{\text{sy}}$ and $r$ with $\rho=-0.50$ and $ P = 8.36 \times 10^{-4}$. We also find a negative correlation between $\nu_{\text{IC}}$ and $r$ with $\rho=-0.41$ and $ P = 0.0076$. The results confirm the consistency of the model and imply that the statistical or stochastic acceleration mechanisms play a major role in \textit{Fermi} TeV BL Lacs jets. Massaro et al. (2004a) pointed out an interesting check that if the log-parabolic electron spectrum is actually related to the energy-dependence statistical acceleration, there should be a linear relation between $r$ and $s$. So we check the correlation among these two parameters, we find that they are significantly correlated with $\rho=0.70$ and $ P = 3.49 \times 10^{-7}$ (see Figure 8). we also do a covariance analysis and a partial correlation analysis to exclude possible covariates (other model parameters) effects. The results of the analyses are listed in Table 4. From the results, we verify that the $r$-$s$ correlation is reliable. The result suggests that the curvature of log-parabolic electron distribution is attributed to the energy-dependence statistical acceleration mechanism working on the emitting electrons. In addition, it is worth noting that the above correlations found are for the source sample (the same as Rani et al. (2011)), and distinguish from Massaro et al. (2004a, 2004b, 2006) that the correlations found are for the spectral evolution of a single source. In Massaro et al. (2004a, 2004b, 2006), the correlations are obtained by dealing with the broadband X-ray spectra of a single source (Mkn 421, Mkn 501) at various epochs.

\begin{figure}
\centering
\includegraphics[width=8cm,height=6cm]{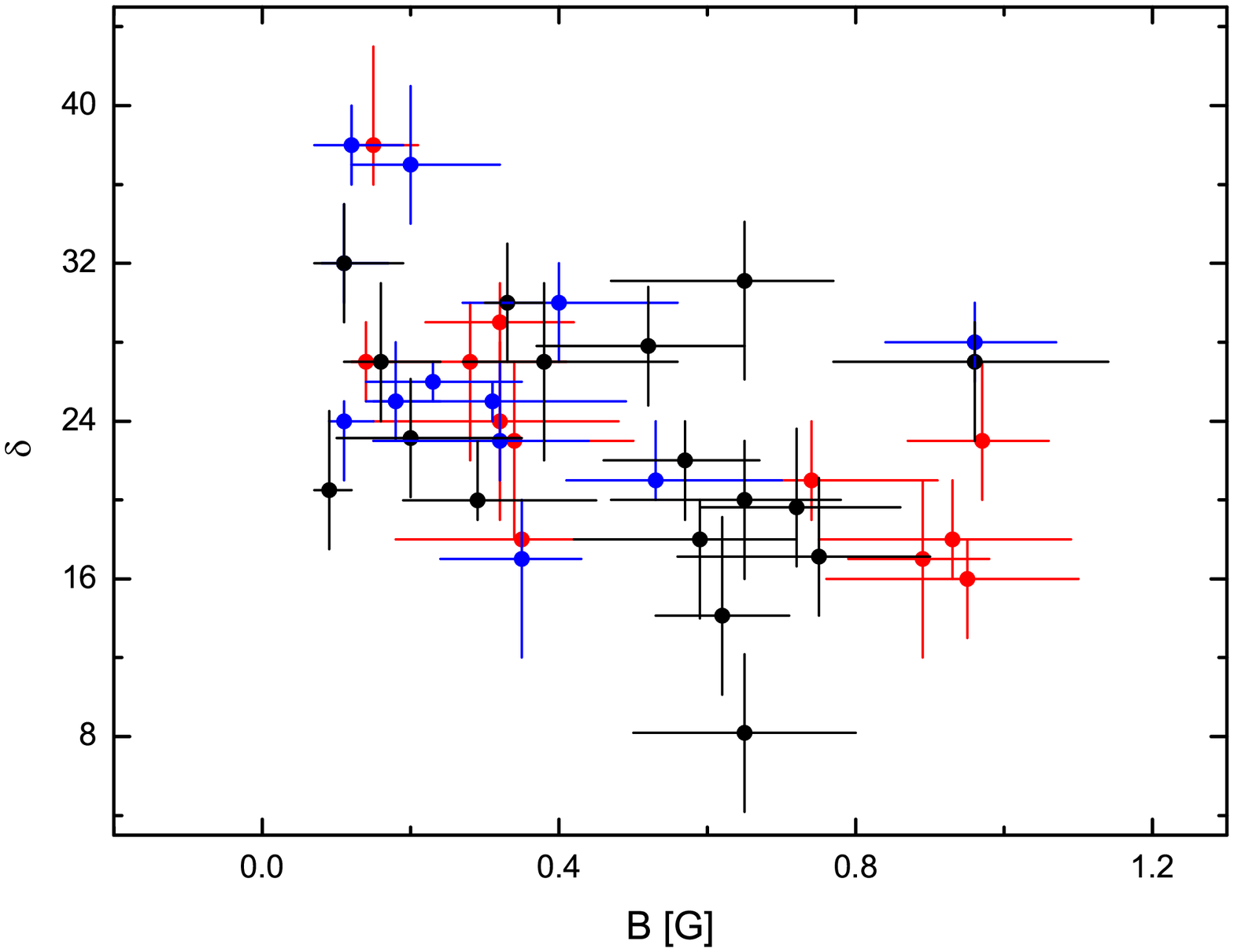}
\caption{The $\delta$ as a function of the $B$. Blue and red circles are for the sources in the high and low states, respectively, and black circles are for the sources with only one SED available.}
\end{figure}
\begin{figure}
\centering
\includegraphics[width=11cm,height=9cm]{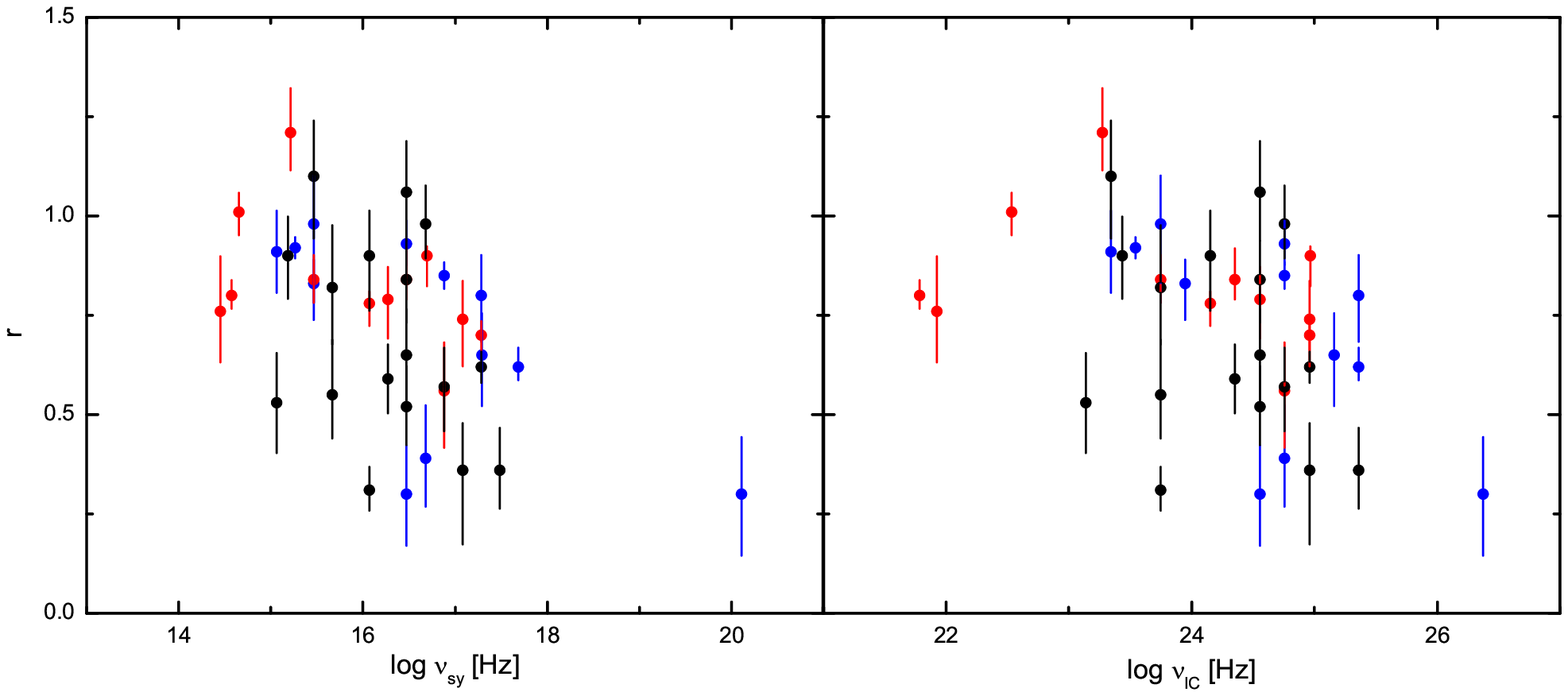}
\caption{The $r$ as a function of the $\nu_{\text{sy}}$ and the $\nu_{\text{IC}}$. The meanings of different symbols are as same as Figure 6.}
\end{figure}
\begin{figure}
\centering
\includegraphics[width=8cm,height=6cm]{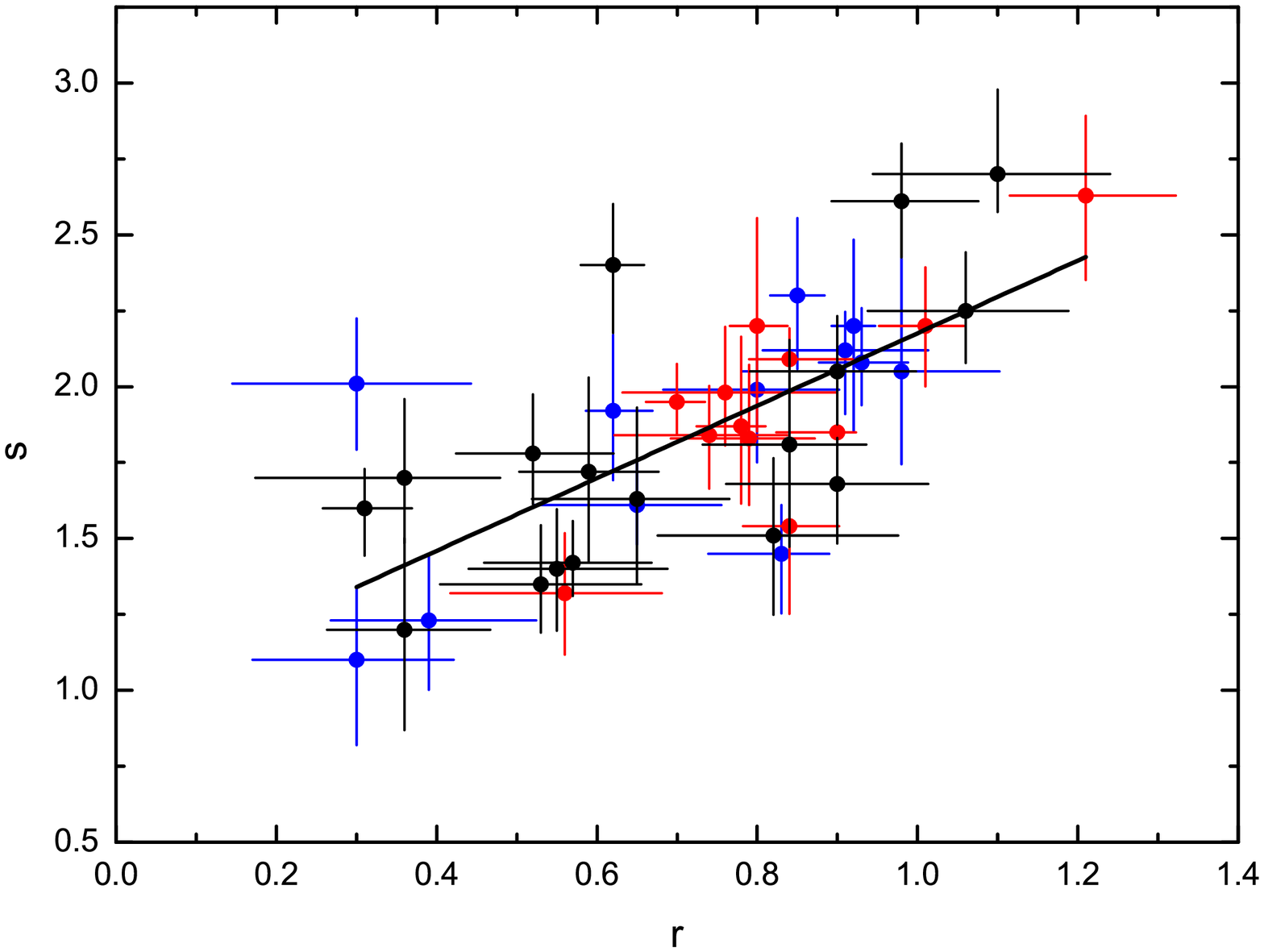}
\caption{The $s$ as a function of the $r$. The meanings of different symbols are as same as Figure 6. The straight line is the best linear fit with a slope, $m=1.19\pm0.19$ and constant, $c = 0.98\pm0.14$ ($y = mx + c$).}
\end{figure}

From the Figure 1, it can be roughly estimated that the ratio between the SSC peak flux to the synchrotron peak flux $(\nu f_\nu)_{\text{IC}}/(\nu f_\nu)_{\text{sy}} < 1$ for most of sources, which implies that the $U_{\text{syn}}/U_{\text{B}} < 1$ and the synchrotron cooling is more important than SSC cooling, where the $U_{\text{B}}=B^{2}/8\pi$ is the magnetic energy density, and the $U_{\text{syn}}$ is synchrotron photon energy density for IC scattering. The effect of radiative cooling will be to move high energy electrons to lower energies, resulting in an increased curvature and a steepened electron distribution at high energies. Because the synchrotron cooling is important than SSC cooling. Therefore, the curvature $r$ of the electron energy distribution may be related to the magnetic field $B$. However, we do not find any correlations between $B$ and $r$ for our sample (See Figure 9, $\rho=0.09$ and $P=0.54$). 
\begin{figure}
\centering
\includegraphics[width=8cm,height=6cm]{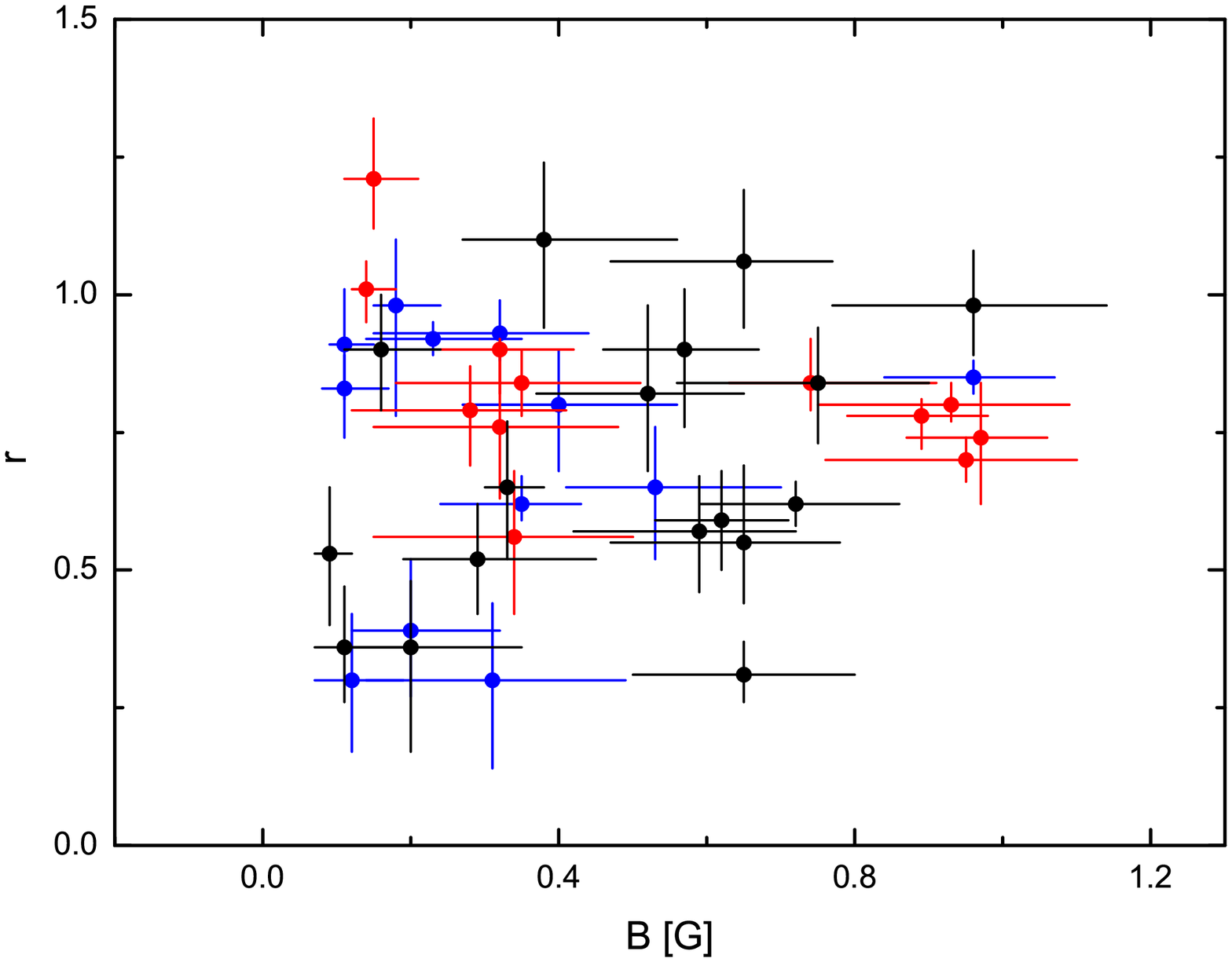}
\caption{The $r$ as a function of the $B$. The meanings of different symbols are as same as Figure 6.}
\end{figure}

\subsection{The physical properties of jet}
Based on the model parameters, we can estimate the jet power and the radiative power. Assuming the jet is heavy (i.e. mainly composed of an electron-proton plasma), one proton per relativistic electron, the protons 'cold' in the comoving frame, and the jet power ($P_{\text{jet}}$) are be carried by relativistic electrons, cold protons, and magnetic field (Celotti \& Ghisellini 2008). The $P_{\text{jet}}$ can be estimated with $P_{\text{jet}} = P_{\text{B}}+P_{\text{e}}+P_{\text{p}}$ in the stationary frame of the host galaxy (e.g., Celotti \& Ghisellini 2008; Ghisellini et al. 2010; Zhang et al. 2012; Yan et al. 2014). The $P_{\text{B}}$ is Poynting flux power, the $P_{\text{e}}$ and the $P_{\text{p}}$ are powers of relativistic electrons and protons, respectively, which are calculated with
\begin{equation}
P_{i} = \pi R^{2} \Gamma^{2} c U_{\text{i}}
\end{equation}
in the stationary frame of the host galaxy (Celotti \& Fabian 1993; Celotti \& Ghisellini 2008), where the $\Gamma$ is the bulk Lorentz factor (we take $\Gamma = \delta$), and the $U_{\text{i}}$ (i = e, p, B) are the energy densities associated with the emitting electrons $U_{\text{e}}$, protons $U_{\text{p}}$, and magnetic field $U_{B}$ in the comoving frame, respectively. The $U_{\text{e}}$ is calculated with
\begin{equation}
U_{\text{e}} = m_{\text{e}}c^{2}\int \gamma N(\gamma)d\gamma.
\end{equation}
The protons are considered 'cold' in the comoving frame, the $U_{\text{p}}$ is calculated with
\begin{equation}
U_{\text{p}} = m_{\text{p}}c^{2}\int N(\gamma)d\gamma
\end{equation}
by assuming one proton per emitting electron (see Celotti \& Ghisellini 2008). The $U_{B}$ is calculated with 
\begin{equation}
U_{\text{B}} = B^{2}/8\pi
\end{equation}
The radiation power is estimated with the observed luminosity,
\begin{equation}
P_{\text{r}} = \pi R^{2} \Gamma^{2} c U_{\text{r}} = L^{\prime} \frac{\Gamma^{2}}{4} = L\frac{\Gamma^{2}}{4\delta^{4}}\approx L\frac{1}{4\delta^{2}},
\end{equation}
where $L$ is the total observed non-thermal luminosity in the Earth ($L^{\prime}$ is in the comoving frame). Because the total observed non-thermal luminosity mainly comes from the jet, especially for TeV BL Lacs, so we take $L \sim L_{\text{bol}}$ (bolometric luminosity of the jet). We calculate $L_{\text{bol}}$ in the band $10^{10}-10^{27}$ Hz based on our SED fits, i.e., $L_{\text{bol}} = 4 \pi d_{L}^{2} \int_{10^{10} \text{Hz}}^{10^{27} \text{Hz}} F_{obs}(\nu_{obs}) d\nu_{obs}$. The calculated $P_{\text{e}}$, $P_{\text{p}}$, $P_{\text{B}}$, and $P_{\text{r}}$ are listed in Table 3. From our SED modeling results (Figure 1), it can be seen that the GeV data at low energies can be fitted well. And the average spectral index between the X-ray and the $\gamma$-ray of our calculated radiation spectrums $\alpha^{\prime}_{\text{X}\gamma}=1.12$ is similar to the average observed value of the high peaked frequency BL Lacs $\alpha_{\text{X}\gamma}=1.02$ (Fan et al. 2012). Which implies the estimated value of $\gamma_{\min}$ is reasonable. In addition, we give the $P_e$ and $P_p$ as a function of the $\gamma_{\min}$ in Figure 10 (Here, the other physical parameters take the average values of the sample). It can be find that the $\gamma_{\min}$ causes minor effect on the estimate of the $P_e$ and $P_p$, especially in $\gamma_{\min}<100$. Therefore, we would like to believe that the powers we derived here are creditable.
\begin{figure}
\centering
\includegraphics[width=9cm,height=6cm]{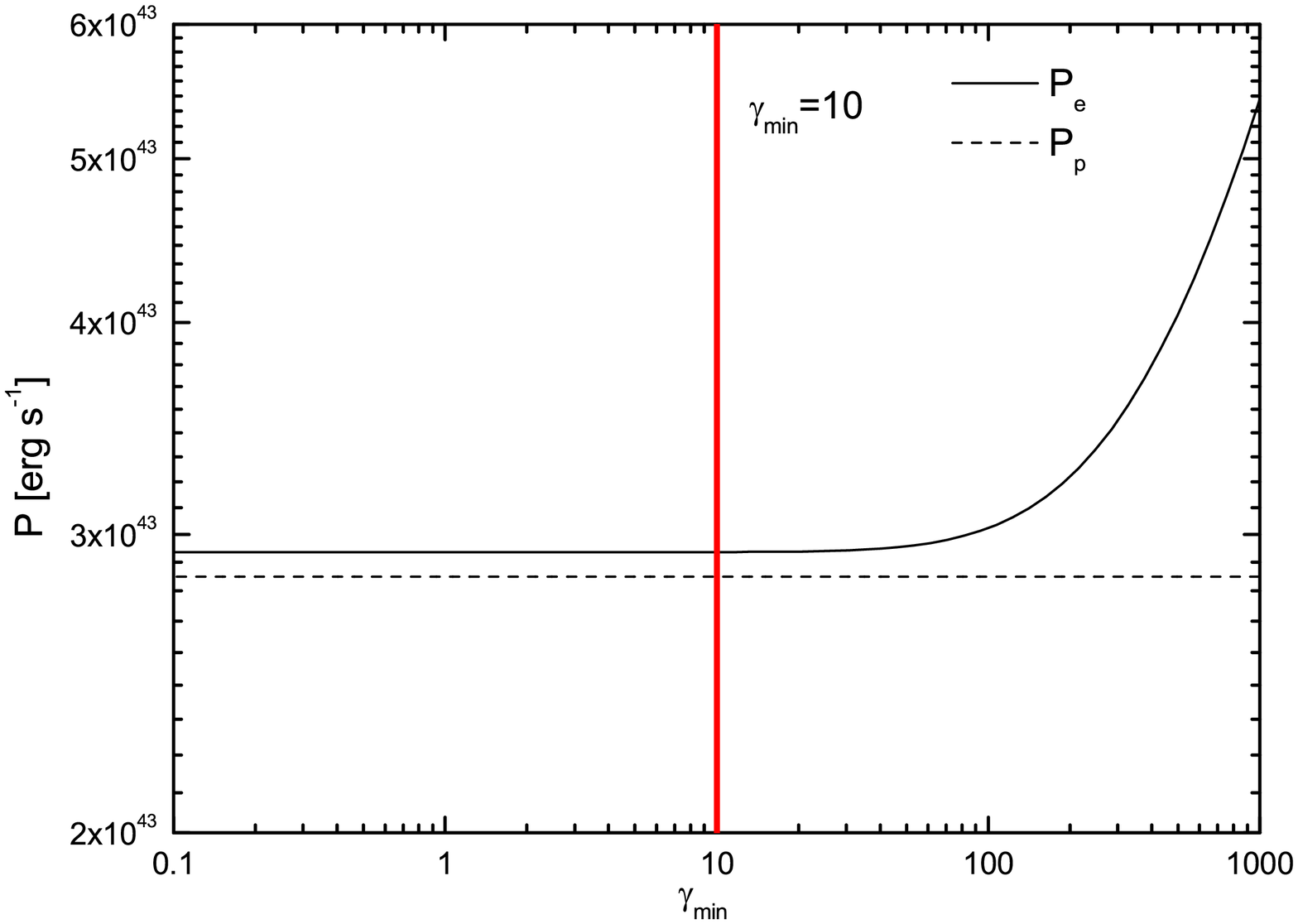}
\caption{The $P_{\text{e}}$, $P_{\text{p}}$ as a function of the $\gamma_{\min}$ when the other physical parameters take  the average values of the sample ($z=0.163, R = 4.28E+15, B = 0.45, \delta = 24, N = 48, r = 0.79, s=1.85, \log \gamma_{\text{0}}= 3.5, \log \gamma_{\text{max}}= 6.2$).}
\end{figure}

In Figure 11, we plot the $P_{\text{e}}$, $P_{\text{p}}$, $P_{\text{B}}$, $P_{\text{jet}}$ as a function of the $P_{\text{r}}$. A size relation $P_{\text{e}} \sim P_{\text{p}} > P_{\text{r}} \gtrsim P_{\text{B}}$ can be found from Figure 11. It should be pointed out that because the definitions of $U_{\text{e}}$, $U_{\text{p}}$, $U_{\text{B}}$ are independent of each other, so the size relation of $P_{\text{e}}$, $P_{\text{p}}$, $P_{\text{B}}$ does not exist functional biases (Celotti \& Ghisellini 2008). The $P_{\text{e}}>P_{\text{B}}$ suggests that the jets are particle dominated. The $P_{\text{e}}\sim P_{\text{p}}$ is consistent with Celotti \& Ghisellini (2008) and means that the mean energy of relativistic electrons approaches $m_{\text{p}}/m_{\text{e}}$ in TeV BL Lacs. The $P_{\text{r}} \gtrsim P_{B}$ implies that Poynting flux cannot account for the radiation power. In fact, it can be found that $P_{\text{r}}/P_{\text{e}} \sim 0.01-1.5$, which indicates that a large fraction of the relativistic electron power will be used to produce the observed radiation in some sources. So, as mentioned in Celotti \& Ghisellini (2008), an additional reservoir of energy is needed to accelerate electrons to high energies for those sources whose the relativistic electron power is comparable to the emitted one. 
\begin{figure}
\centering
\includegraphics[width=10cm,height=9cm]{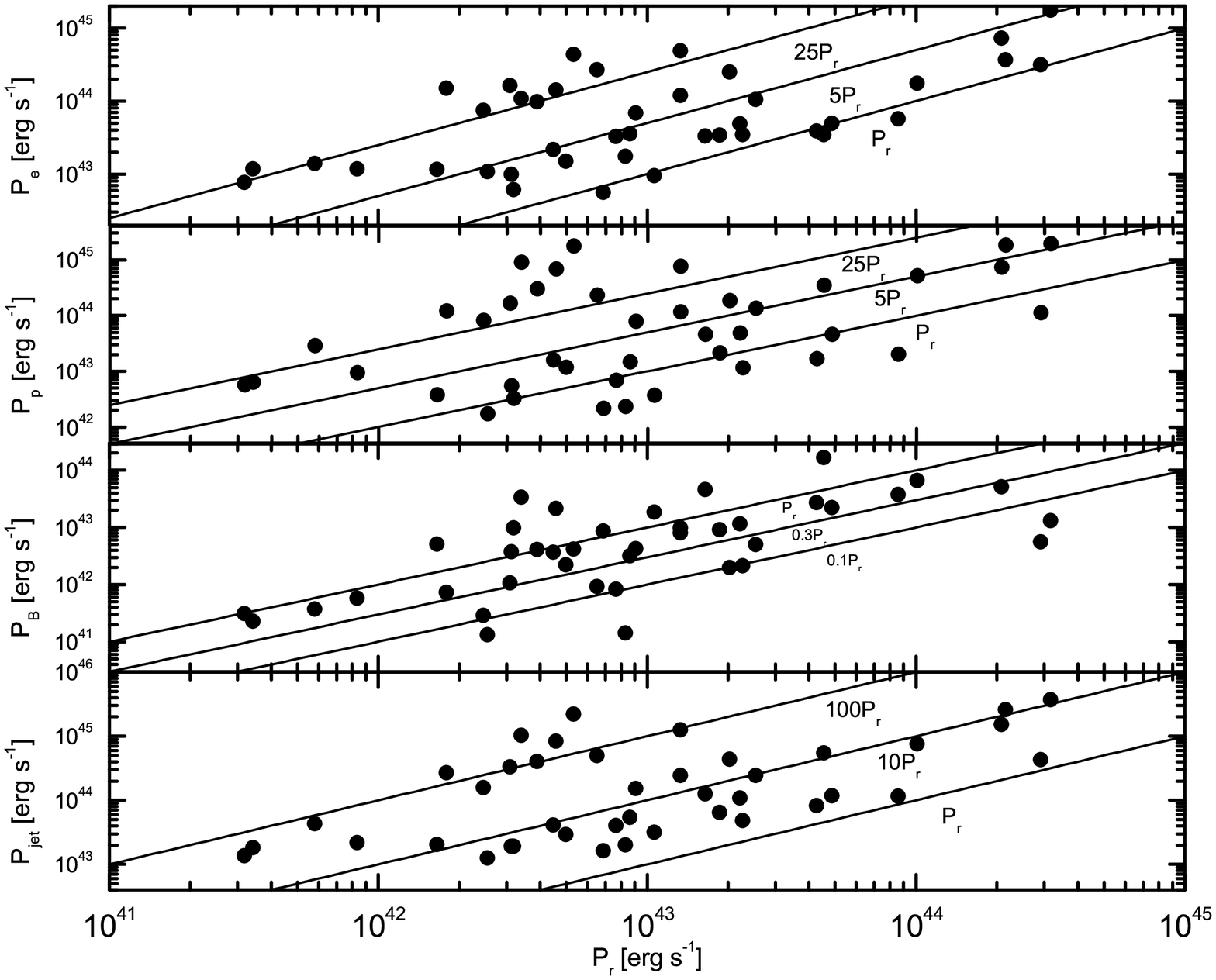}
\caption{The $P_{\text{e}}$, $P_{\text{p}}$, $P_{\text{B}}$, $P_{\text{jet}}$ as a function of the $P_{\text{r}}$.}
\end{figure}

It is generally believed that jet are driven by the accretion process and/or the spin of central black hole. Davis \& Laor (2011) suggested that the black hole mass would be an essential factor for the jet radiation efficiency and jet power. We investigate the relation of $P_{\text{jet}}$ to the $M$ with a sub-sample of 18 sources in our sample. The black hole mass $M$ of these sources are collected from literature and reported (see Table 3), when more than one black hole masses are got, we use average value. The $P_{\text{jet}}$ as a function of the $M$ is shown in Figure 12. We do not find significant correlations between the two parameters in high data ($\rho=-0.56$, $P=0.09$) and in low state data ($\rho=-0.48$, $P=0.16$). The result suggests that the jet power weakly depends on the black hole mass for \textit{Fermi} TeV BL Lac jets.
\begin{figure}
\centering
\includegraphics[width=8cm,height=6cm]{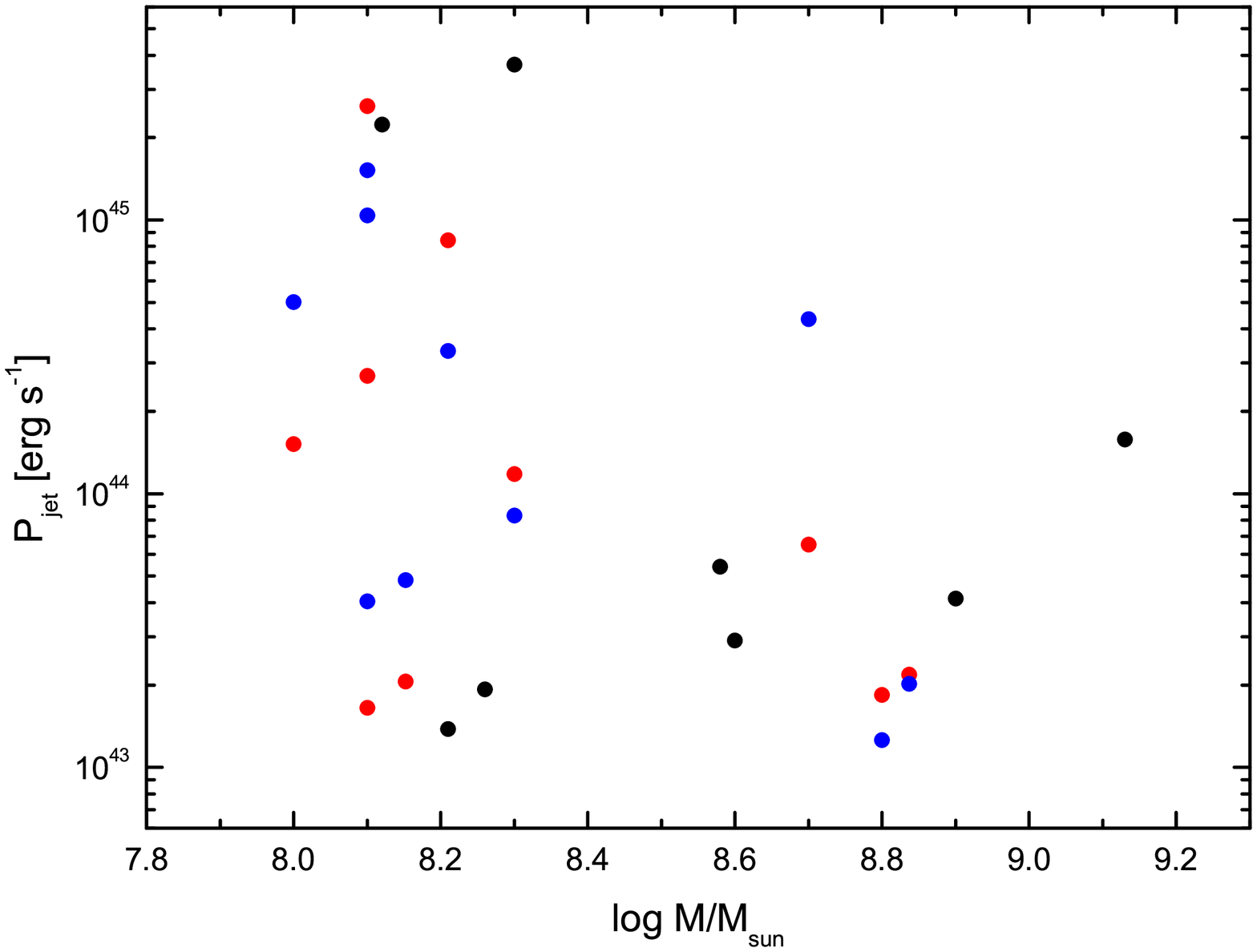}
\caption{The $P_{\text{jet}}$ as a function of the black hole mass $M$. The meanings of different symbols are as same as Figure 6.}
\end{figure}

In Figure 13, we show that there is an anti-correlation between $\gamma_{\text{peak}}$ and $P_{\text{jet}}$, i.e., $P_{\text{jet}} \propto \gamma_{\text{peak}}^{-1.02\pm0.19}$ with $\rho=-0.647$ and $P=4.9\times10^{-6}$, where $\gamma_{\text{peak}}$ is the electrons energy at the peaks of the SED. For the log-parabolic electron energy distribution here adopted, the $\gamma_{\text{peak}}$ = $\gamma_{\text{3p}}$ (the peak energy of the distribution $\gamma^3 N(\gamma)$) and are calculated as given by Paggi et al. (2009b),
\begin{equation}
\gamma_{\text{peak}} = \gamma_{\text{3p}} = \gamma_{\text{p}} \times 10^{1/2r}.
\end{equation}
where $\gamma_{\text{p}} = (\int \gamma^{2}N(\gamma)d\gamma/\int N(\gamma)d\gamma)^{1/2}$. The anti-correlation between $\gamma_{\text{peak}}$ and $P_{\text{jet}}$ is consistent with the prediction of the blazar sequence (Fossati et al. 1998; Ghisellini et al. 1998; Celotti \& Ghisellini (2008); Ghisellini \& Tavecchio 2008) and is usually explained as that the radiative cooling is stronger in more powerful blazars.
\begin{figure}
\centering
\includegraphics[width=8cm,height=6cm]{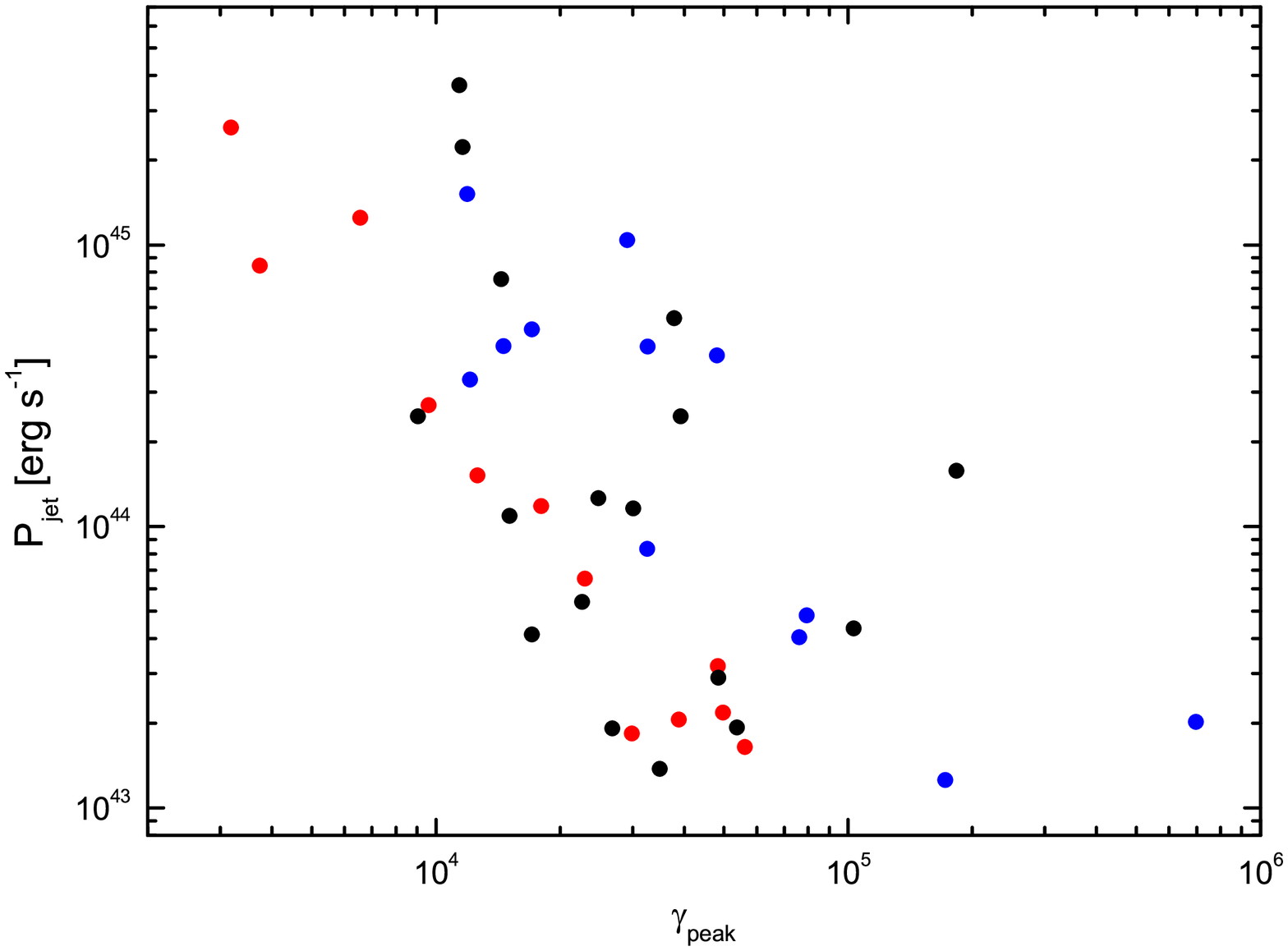}
\caption{The $P_{\text{jet}}$ as a function of the $\gamma_{\text{peak}}$. The meanings of different symbols are as same as Figure 6.}
\end{figure}

\subsection{The cause of flux variation}
\begin{figure}
\centering
\includegraphics[width=8cm,height=6cm]{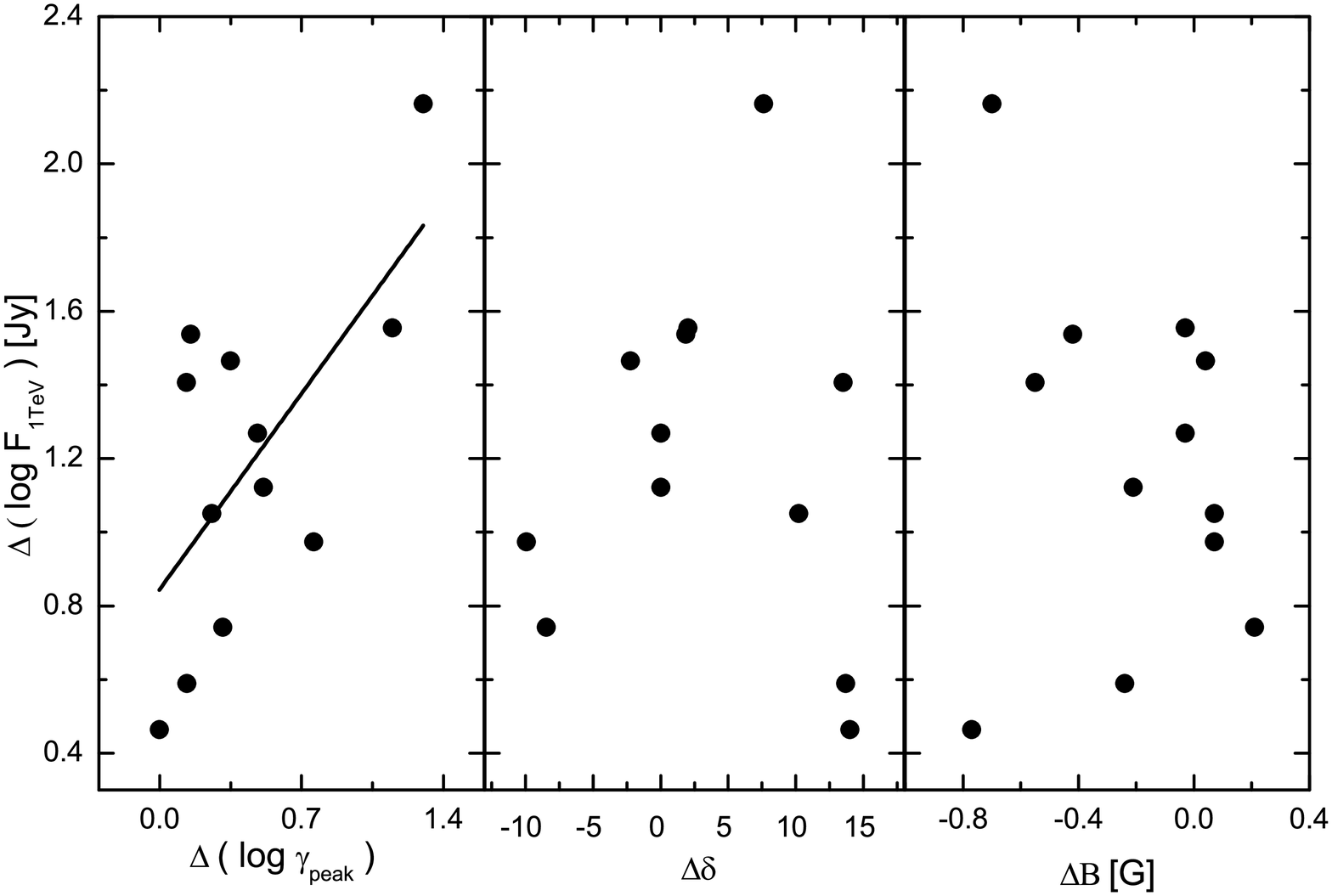}
\caption{The changes in the flux density at 1 TeV ($F_{\text{1TeV}}$) as a function of the changes in $\gamma_{\text{peak}}$, $\delta$, and $B$. The straight line is the best linear fit with a slope, $m=0.75\pm0.27$ and constant, $c = 0.84\pm0.16$ ($y = mx + c$).}
\end{figure}
The SEDs of 14 sources in our sample are obtained the low and high TeV states. There are several possibilities that could be invoked to explain the flux variation. 1) It is well known that the electron energy distribution is influenced by several processes, such as acceleration, escape and cooling. It is possible that the evolution of the electron energy distribution leads to the flux variation. 2) Another reason to explain the flux variation is the change of the Doppler boosting factor, presumably due to a change either the bulk velocity or the viewing angle. Raiteri et al. (2010) found that only geometrical (Doppler factor) changes are capable of explaining flux variation for BL Lacs. 3) In addition, the emission dissipation at different location in the jet also may cause the flux variation. Since the value of the magnetic field $B$ is related to the location of the dissipation region. In this case, we expect to find that the change in $B$ value is related to the flux variation. 

we plot the changes in the flux density at 1 TeV ($F_{\text{1TeV}}$) as a function of the changes in $\gamma_{\text{peak}}$, $\delta$, $B$ in Figure 14. We find a correlation between the changes in the flux density at 1 TeV and the changes in the $\gamma_{\text{peak}}$ ($\rho=0.66$, $P=0.02$), but no statistically correlations between the changes in the flux density at 1 TeV and the changes in $\delta$ ($\rho=-0.07$, $P=0.83$) and $B$ ($\rho=-0.20$, $P=0.54$) are found, which suggests the change/evolution of electron energy distribution may be mainly responsible for the flux variation. Our result is similar to Zhang et al. (2012) and supports the research results of single TeV BL Lac, e.g., Zheng et al. (2011a, 2011b) investigated the rapid TeV flare in Mkn 501 and multi-wavelength variability in PKS 2155-304, their result showed that the flux variation is caused by the evolution of electron energy distribution. Paggi et al. (2009b) investigated S5 0716+714, Mkn 501, Mkn 421 and found that the flares are directly related to acceleration of the emitting electrons.

\section{CONCLUSIONS}
In this work, we have modeled the quasi-simultaneous broadband SEDs of 29 \textit{Fermi} TeV BL Lac objects by using a one-zone leptonic SSC model with the log-parabolic electron energy distribution. We obtain the best-fit model parameters by MCMC sampling method. Then, we systematically investigate the physical properties of \textit{Fermi} TeV BL Lac jets. Our main results are the following: (i) There is a negative correlation between $B$ and $\delta$ for our source sample, which suggests that $B$ and $\delta$ are dependent on each other mainly in Thomson regime. (ii) There are negative correlations between $\nu_{\text{sy}}$ and $r$, $\nu_{\text{IC}}$ and $r$ for our source sample, which confirms the consistency of the model and is a signature of the energy-dependence statistical acceleration or the stochastic acceleration. We check the correlation between $r$ and $s$, and find that they are significantly correlated. The result suggests that the curvature of log-parabolic electron energy distribution is attributed to the energy-dependence statistical acceleration mechanism working on the emitting electrons in \textit{Fermi} TeV BL Lacs. We do not find any correlations between $B$ and $r$, which may be because the curvature of the electron energy distribution is related to the particle acceleration rather than to the radiative cooling process. However, it should be pointed out that because we use a "static" SSC code and do not consider the evolution of the electron energy distribution, so the effects of radiation cooling on the electron energy distribution are not really addressed in here. 
(iii) By assuming one proton per relativistic electron and the jet power are be carried by relativistic electrons, cold protons, magnetic field (Celotti \& Ghisellini 2008). We calculate the jet power in different forms and the radiative power in the stationary frame of the host galaxy. We find a size relation $P_{\text{e}} \sim P_{\text{p}} > P_{\text{r}} \gtrsim P_{\text{B}}$. The $P_{\text{e}}>P_{\text{B}}$ suggests that the jets are particle dominated in \textit{Fermi} TeV BL Lacs. The $P_{\text{e}}\sim P_{\text{p}}$ is consistent with Celotti \& Ghisellini (2008) and means that the mean energy of relativistic electrons approaches $m_{\text{p}}/m_{\text{e}}$ in \textit{Fermi} TeV BL Lacs. Then, we explore the correlation between $P_{jet}$ and black hole masses with a sub-sample of 18 sources in our sample. There are not significant correlations whether in high or low state, which suggests that the jet power weakly depends on the black hole mass for \textit{Fermi} TeV BL Lac jets. And we find the $P_{\text{jet}} \propto \gamma_{\text{peak}}^{-1.02\pm0.19}$, which is consistent with the prediction of the blazar sequence. (iv) At last, we explore the cause of flux variation. We only find a correlation between the changes in the flux density at 1 TeV and the changes in the $\gamma_{\text{peak}}$, which suggests the change/evolution of electron energy distribution may be mainly responsible for the flux variation in our sample.

\section*{Acknowledgments}
We sincerely thank anonymous referee for valuable comments and suggestions. We are very grateful to the Science Foundation of Yunnan Province of China (2012FB140, 2010CD046). This work is supported by the National Nature Science Foundation of China (11063004, 11163007, U1231203), and the High-Energy Astrophysics Science and Technology Innovation Team of Yunnan Higher School and Yunnan Gravitation Theory Innovation Team (2011c1). This research has made use of the NASA/IPAC Extragalactic Database (NED), that is operated by Jet Propulsion Laboratory, California Institute of Technology, under contract with the National Aeronautics and Space Administration.

\begin{table*}
\centering
\begin{minipage}{250mm}
\caption{The time of quasi-simultaneous observation, the observed telescopes, and the references for each source}
\scriptsize
\begin{tabular}{@{}ccllllllll@{}}
\hline\hline
Name	&	State	&	Time   & 	The observed telescopes	&	Ref	\\
{(1)} & {(2)} & {(3)} & {(4)} & {(5)}\\
\hline
1ES 1101-232	&	H	&	2006-May	&	\textit{Suzaku}, H.E.S.S.	&	Reimer et al. 2008	\\
	&	L	&	2004-Jun to 2005-Mar	&	XMM-Newton, \textit{RXTE}, H.E.S.S.	&	Aharonian et al. 2007	\\
Mkn421	&	H	&	2003-Feb to 2004-Jun	&	FLWO, \textit{RXTE}, Whipple	&	Blazejowski et al. 2005	\\
	&	L	&	2009-Jan to 2009-Jun	&	\textit{Swift}-UVOT, \textit{Swift}-XRT, \textit{Swift}-BAT, \textit{Fermi}-LAT, MAGIC	&	Abdo et al. 2011a	\\
Mkn501	&	H	&	1997-Apr	&	\textit{BeppoSAX}, CAT	&	Tavecchio et al. 2001	\\
	&	L	&	2006-Jul	&	KVA, \textit{Suzaku}, MAGIC	&	Anderhub et al. 2009a	\\
1ES 1959+650	&	H	&	2002-May to 2002-Jun	&	Boltwood Observatory, Abastumani Observatory, \textit{RXTE}, HEGRA	&	Krawczynski et al. 2004	\\
	&	L	&	2006-May	&	\textit{Swift}-UVOT, \textit{Suzaku}, MAGIC	&	Tagliaferri et al. 2008	\\
PKS 2005-489	&	H	&	2009-May to 2009 July	&	\textit{Swift}-UVOT, \textit{RXTE}, \textit{Swift}-XRT, H.E.S.S.	&	Kaufmann et al. 2010	\\
	&	L	&	2004-Oct	&	 XMM-Newton	&	H.E.S.S. Collaboration et al. 2010a	\\
PKS 2155-304	&	H	&	2006-Jul	&	Bronberg Observatory, \textit{RXTE}, Chandra, H.E.S.S.	&	Aharonian et al. 2009a	\\
	&	L	&	2008-Aug to 2008-Sep	&	ATOM, \textit{RXTE}, H.E.S.S.	&	Aharonian et al. 2009b	\\
1ES 2344+514	&	H	&	2007-Dec	&	\textit{Swift}-UVOT, \textit{Swift}-XRT, VERITAS	&	Acciari et al. 2011	\\
	&	L	&	   2005-Apr to 2006-Jan	&	KVA, \textit{Swift}-XRT, MAGIC	&	Albert et al. 2007a; Tramacere et al. 2007b	\\
 W Comae	&	H	&	2008-Jun	&	\textit{Swift}-UVOT, \textit{Swift}-XRT, VERITAS	&	Acciari et al. 2009	\\
	&	L	&	2008-May (low state)	&	AAVSO, \textit{Swift}-UVOT, \textit{Swift}-XRT, \textit{Fermi}-LAT$^{\ast}$, VERITAS	&	  Acciari et al. 2008; Tavecchio et al. 2010	\\
S5 0716+714	&	H	&	2008-Apr	&	KVA, \textit{Swift}-XRT, MAGIC	&	Anderhub et al. 2009b	\\
	&	L	&	2008-Nov	&	\textit{Swift}-UVOT, \textit{Swift}-XRT	&	Tavecchio et al. 2010	\\
 BL Lacertae	&	H	&	2005-Aug to 2005-Dec (low state)	&	KVA, EGRET$^{\ast}$, MAGIC	&	Albert et al. 2007b	\\
	&	L	&	2008-Aug to 2008-Oct	&	Effelsberg Observatory, \textit{Swift}-UVOT, GASP-WEBT Collaboration, \textit{Swift}-XRT, \textit{Fermi}-LAT	&	Abdo et al. 2010a	\\
 1ES 1011+496 	&	H	&	2008-May (high state)	&	\textit{Swift}-UVOT, \textit{Swift}-XRT, MAGIC$^{\ast}$	&	Albert et al. 2007c; Tavecchio et al. 2010	\\
	&	L	&	2008-Aug to 2008-Oct	&	OVRO, \textit{Swift}-UVOT, \textit{Swift}-XRT, \textit{Fermi}-LAT	&	Abdo et al. 2010a	\\
 MAGIC J2001+435	&	H	&	 2010-Jul	&	\textit{Swift}-UVOT, \textit{Swift}-XRT, MAGIC	&	Aleksi\'{c} et al. 2014b	\\
	&	L	&	2010-Jul to 2010-Sep (low state)	&	OVRO, \textit{Swift}-UVOT, \textit{Swift}-XRT, \textit{Fermi}-LAT	&	Aleksi\'{c} et al. 2014b	\\
Mkn180	&		&	2008-May (low state)	&	\textit{Swift}-UVOT, \textit{Swift}-XRT, MAGIC$^{\ast}$	&	R\"{u}gamer et al. 2011	\\
 3C 66A	&		&	 2008-Oct 	&	MDM, \textit{Swift}-UVOT, \textit{Swift}-XRT, \textit{Fermi}-LAT, VERITAS	&	Abdo et al. 2011b	\\
PKS 1424+240	&		&	2009-Feb to 2009-Jun 	&	MDM, \textit{Swift}-UVOT, \textit{Swift}-XRT, \textit{Fermi}-LAT, VERITAS	&	Acciari et al. 2010a	\\
1ES 1215+303	&		&	2011-Jan to 2011-Feb	&	Mets\"{a}hovi, KVA, \textit{Swift}-UVOT, \textit{Swift}-XRT, \textit{Fermi}-LAT, MAGIC	&	Aleksi\'{c} et al. 2012a	\\
1RXS J003334.6-192130	&		&	2008-May to 2008-Nov	&	\textit{Swift}-UVOT, \textit{Swift}-XRT, \textit{Fermi}-LAT	&	Abdo et al. 2010a	\\
RGB J0152+017	&		&	2007-Nov	&	\textit{RXTE}, \textit{Swift}-XRT, H.E.S.S.	&	Aharonian et al. 2008	\\
 1ES 0414+009 	&		&	2010-Jul to 2011-Feb	&	MDM, \textit{Swift}-XRT, \textit{Fermi}-LAT, VERITAS	&	Aliu et al. 2012	\\
 PKS 0447-439	&		&	2009-Nov to 2010 Jan	&	\textit{Swift}-UVOT, \textit{Swift}-XRT, \textit{Fermi}-LAT, H.E.S.S.	&	Prandini et al. 2012	\\
RGB J0710+591	&		&	2008-Dec to 2009-Mar	&	MDM, \textit{Swift}-UVOT, \textit{Swift}-XRT, VERITAS	&	Acciari et al. 2010b	\\
1ES 0806+524 	&		&	2007-Nov to 2008-Apr	&	\textit{Swift}-UVOT, \textit{Swift}-XRT, VERITAS	&	Acciari et al. 2009b	\\
MS 1221.8+2452	&		&	2013-Mar to 2013-Apr (low state)	&	OVRO, \textit{Swift}-UVOT, \textit{Swift}-XRT, MAGIC$^{\ast}$, VERITAS$^{\ast}$	&	Singh et al. 2014	\\
H 1426+428	&		&	low state	&	\textit{BeppoSAX}$^{\ast}$, INTEGRAL$^{\ast}$, CAT$^{\ast}$, HEGRA$^{\ast}$	&	Wolter et al. 2008	\\
PG 1553+113	&		&	2006-Apr to 2006 Jul (low state)	&	\textit{Suzaku}, \textit{Fermi}-LAT$^{\ast}$, MAGIC, H.E.S.S.	&	Abdo et al. 2010b	\\
 B3 2247+381 	&		&	2010-Sep to 2010-Oct (low state)	&	Plank$^{\ast}$, KVA, \textit{Swift}-XRT, \textit{Fermi}-LAT$^{\ast}$, MAGIC	&	Aleksi\'{c} et al. 2012b	\\
1ES 1741+196	&		&	2010-Feb to 2010-Apr	&	ATCA, Plank, \textit{Swift}-UVOT, \textit{Swift}-XRT, \textit{Fermi}-LAT 	&	Giommi et al. 2012	\\
1ES 1218+30.4	&		&	2005-Jan to 2005-March	&	KVA, \textit{Swift}-XRT, MAGIC 	&	R\"{u}ger et al. 2010	\\
H 2356-309	&		&	2005-June	&	XMM-Newton, H.E.S.S.	&	H.E.S.S. Collaboration et al. 2010b	\\
\hline
\end{tabular}
\end{minipage}
\flushleft{\textbf{Notes.} Column (2) is the state of each source in TeV band. "H" indicating "high state" and "L" indicating "low state". Column (3) is the time of quasi-simultaneous observation, the content in the bracket is the state of the source in the time of quasi-simultaneous observation. In here, H 1426+428 has no quasi-simultaneous observation, we use its SED data that it is in low state. Column (4) is the observed telescopes, the superscript  '$^{\ast}$'} represents that the data of observed telescope are not in the time of quasi-simultaneous observation, but the state of these data consistent with that in the time of quasi-simultaneous observation. Column (5) is the references of the quasi-simultaneous broadband SEDs.
\end{table*}

\begin{table*}
\centering
\begin{minipage}{250mm}
\caption{The best-fit model parameters and the flux density at 1 TeV}
\scriptsize
\begin{tabular}{@{}cccrrrrrrrrrrrrrrrrrrrrrrrrrrrr@{}}
\hline\hline
Name & State & $z$ & $R$~(cm) & $B$ ~(G) & $\delta$ & $r$ & $s$ & $N$ & $\log \gamma_{0}$ & $\log \gamma_{\max}$ & $F_{\text{1TeV}}$~(Jy) \\
{(1)} & {(2)} & {(3)} & {(4)} & {(5)} & {(6)} & {(7)} & {(8)} & {(9)} & {(10)}& {(11)} & {(12)} \\
\hline
1ES 1101-232	&	H	&	0.186	&	4.47E+15	&	0.20 	$^{+	0.12 	}	_{-	0.08 	}$	&	37 	$^{+	4 	}	_{-	3 	}$	&	0.39 	$^{+	0.13 	}	_{-	0.12 	}$	&	1.23 	$^{+	0.21 	}	_{-	0.23 	}$	&	78 	$^{+	3 	}	_{-	3 	}$	&	2.4 	$^{+	0.2 	}	_{-	0.3 	}$	&	6.2 	$^{+	0.2 	}	_{-	0.4 	}$	&	1.1E-15	\\
	&	L	&		&	3.16E+15	&	0.97 	$^{+	0.09 	}	_{-	0.10 	}$	&	23 	$^{+	4 	}	_{-	3 	}$	&	0.74 	$^{+	0.10 	}	_{-	0.12 	}$	&	1.84 	$^{+	0.16 	}	_{-	0.18 	}$	&	5 	$^{+	2 	}	_{-	4 	}$	&	3.9 	$^{+	0.2 	}	_{-	0.2 	}$	&	6.5 	$^{+	0.3 	}	_{-	0.2 	}$	&	3.8E-16	\\
Mkn421	&	H	&	0.030021	&	2.19E+15	&	0.53 	$^{+	0.17 	}	_{-	0.12 	}$	&	21 	$^{+	3 	}	_{-	1 	}$	&	0.65 	$^{+	0.11 	}	_{-	0.13 	}$	&	1.61 	$^{+	0.14 	}	_{-	0.13 	}$	&	40 	$^{+	3 	}	_{-	3 	}$	&	3.8 	$^{+	0.3 	}	_{-	0.2 	}$	&	6.8 	$^{+	0.6 	}	_{-	0.7 	}$	&	1.6E-13	\\
	&	L	&		&	3.98E+15	&	0.32 	$^{+	0.10 	}	_{-	0.10 	}$	&	29 	$^{+	2 	}	_{-	2 	}$	&	0.90 	$^{+	0.02 	}	_{-	0.08 	}$	&	1.85 	$^{+	0.28 	}	_{-	0.21 	}$	&	2 	$^{+	4 	}	_{-	1 	}$	&	4.0 	$^{+	0.2 	}	_{-	0.2 	}$	&	6.7 	$^{+	0.5 	}	_{-	0.6 	}$	&	2.9E-14	\\
Mkn501	&	H	&	0.033663	&	7.94E+14	&	0.31 	$^{+	0.18 	}	_{-	0.17 	}$	&	25 	$^{+	1 	}	_{-	1 	}$	&	0.30 	$^{+	0.14 	}	_{-	0.16 	}$	&	2.01 	$^{+	0.21 	}	_{-	0.22 	}$	&	42 	$^{+	2 	}	_{-	5 	}$	&	5.0 	$^{+	0.3 	}	_{-	0.4 	}$	&	6.7 	$^{+	0.7 	}	_{-	0.6 	}$	&	1.4E-13	\\
	&	L	&		&	1.58E+15	&	0.34 	$^{+	0.16 	}	_{-	0.19 	}$	&	23 	$^{+	4 	}	_{-	5 	}$	&	0.56 	$^{+	0.12 	}	_{-	0.14 	}$	&	1.32 	$^{+	0.20 	}	_{-	0.20 	}$	&	50 	$^{+	5 	}	_{-	3 	}$	&	3.2 	$^{+	0.1 	}	_{-	0.1 	}$	&	6.0 	$^{+	0.7 	}	_{-	0.6 	}$	&	3.9E-15	\\
1ES 1959+650	&	H	&	0.047	&	1.26E+15	&	0.40 	$^{+	0.16 	}	_{-	0.13 	}$	&	30 	$^{+	2 	}	_{-	3 	}$	&	0.80 	$^{+	0.10 	}	_{-	0.12 	}$	&	1.99 	$^{+	0.18 	}	_{-	0.24 	}$	&	35 	$^{+	3 	}	_{-	2 	}$	&	4.3 	$^{+	0.3 	}	_{-	0.3 	}$	&	6.3 	$^{+	0.4 	}	_{-	0.4 	}$	&	5.1E-14	\\
	&	L	&		&	3.16E+15	&	0.95 	$^{+	0.15 	}	_{-	0.19 	}$	&	16 	$^{+	2 	}	_{-	3 	}$	&	0.70 	$^{+	0.04 	}	_{-	0.04 	}$	&	1.95 	$^{+	0.12 	}	_{-	0.11 	}$	&	6 	$^{+	1 	}	_{-	3 	}$	&	4.0 	$^{+	0.1 	}	_{-	0.2 	}$	&	6.3 	$^{+	0.2 	}	_{-	0.1 	}$	&	2.0E-15	\\
PKS 2005-489	&	H	&	0.071	&	2.09E+16	&	0.12 	$^{+	0.07 	}	_{-	0.05 	}$	&	38 	$^{+	2 	}	_{-	2 	}$	&	0.30 	$^{+	0.12 	}	_{-	0.13 	}$	&	1.10 	$^{+	0.24 	}	_{-	0.28 	}$	&	10 	$^{+	2 	}	_{-	4 	}$	&	1.5 	$^{+	0.2 	}	_{-	0.2 	}$	&	6.3 	$^{+	0.2 	}	_{-	0.2 	}$	&	2.6E-15	\\
	&	L	&		&	2.45E+15	&	0.15 	$^{+	0.06 	}	_{-	0.04 	}$	&	38 	$^{+	5 	}	_{-	2 	}$	&	1.21 	$^{+	0.11 	}	_{-	0.09 	}$	&	2.63 	$^{+	0.26 	}	_{-	0.28 	}$	&	97 	$^{+	7 	}	_{-	5 	}$	&	3.8 	$^{+	0.2 	}	_{-	0.2 	}$	&	6.5 	$^{+	0.2 	}	_{-	0.3 	}$	&	1.4E-16	\\
PKS 2155-304	&	H	&	0.116	&	5.25E+15	&	0.32 	$^{+	0.12 	}	_{-	0.17 	}$	&	23 	$^{+	4 	}	_{-	2 	}$	&	0.93 	$^{+	0.06 	}	_{-	0.05 	}$	&	2.08 	$^{+	0.18 	}	_{-	0.14 	}$	&	55 	$^{+	5 	}	_{-	4 	}$	&	4.0 	$^{+	0.4 	}	_{-	0.3 	}$	&	5.8 	$^{+	0.6 	}	_{-	0.4 	}$	&	2.9E-13	\\
	&	L	&		&	3.16E+15	&	0.74 	$^{+	0.17 	}	_{-	0.11 	}$	&	21 	$^{+	3 	}	_{-	2 	}$	&	0.84 	$^{+	0.08 	}	_{-	0.05 	}$	&	2.09 	$^{+	0.10 	}	_{-	0.11 	}$	&	34 	$^{+	3 	}	_{-	2 	}$	&	3.8 	$^{+	0.2 	}	_{-	0.2 	}$	&	5.9 	$^{+	0.5 	}	_{-	0.6 	}$	&	8.4E-15	\\
1ES 2344+514	&	H	&	0.044	&	1.00E+15	&	0.35 	$^{+	0.08 	}	_{-	0.11 	}$	&	17 	$^{+	3 	}	_{-	5 	}$	&	0.62 	$^{+	0.05 	}	_{-	0.03 	}$	&	1.92 	$^{+	0.25 	}	_{-	0.23 	}$	&	42 	$^{+	4 	}	_{-	2 	}$	&	4.4 	$^{+	0.2 	}	_{-	0.2 	}$	&	5.9 	$^{+	0.6 	}	_{-	0.7 	}$	&	1.6E-14	\\
	&	L	&		&	1.02E+15	&	0.28 	$^{+	0.13 	}	_{-	0.16 	}$	&	27 	$^{+	3 	}	_{-	5 	}$	&	0.79 	$^{+	0.08 	}	_{-	0.10 	}$	&	1.83 	$^{+	0.24 	}	_{-	0.22 	}$	&	58 	$^{+	4 	}	_{-	4 	}$	&	3.7 	$^{+	0.3 	}	_{-	0.3 	}$	&	5.5 	$^{+	0.4 	}	_{-	0.2 	}$	&	1.7E-15	\\
 W Comae	&	H	&	0.102	&	4.47E+15	&	0.11 	$^{+	0.06 	}	_{-	0.03 	}$	&	32 	$^{+	3 	}	_{-	2 	}$	&	0.83 	$^{+	0.06 	}	_{-	0.09 	}$	&	1.45 	$^{+	0.16 	}	_{-	0.20 	}$	&	80 	$^{+	3 	}	_{-	4 	}$	&	3.3 	$^{+	0.3 	}	_{-	0.4 	}$	&	6.4 	$^{+	0.5 	}	_{-	0.3 	}$	&	6.6E-15	\\
	&	L	&		&	5.25E+15	&	0.35 	$^{+	0.16 	}	_{-	0.17 	}$	&	18 	$^{+	1 	}	_{-	3 	}$	&	0.84 	$^{+	0.06 	}	_{-	0.06 	}$	&	1.54 	$^{+	0.19 	}	_{-	0.29 	}$	&	60 	$^{+	3 	}	_{-	4 	}$	&	3.2 	$^{+	0.3 	}	_{-	0.3 	}$	&	6.4 	$^{+	0.5 	}	_{-	0.3 	}$	&	1.7E-15	\\
S5 0716+714	&	H	&	0.3	&	2.00E+16	&	0.23 	$^{+	0.12 	}	_{-	0.09 	}$	&	26 	$^{+	1 	}	_{-	1 	}$	&	0.92 	$^{+	0.03 	}	_{-	0.03 	}$	&	2.20 	$^{+	0.28 	}	_{-	0.35 	}$	&	20 	$^{+	3 	}	_{-	3 	}$	&	3.6 	$^{+	0.2 	}	_{-	0.2 	}$	&	5.8 	$^{+	0.5 	}	_{-	0.3 	}$	&	4.7E-16	\\
	&	L	&		&	2.00E+16	&	0.93 	$^{+	0.16 	}	_{-	0.18 	}$	&	18 	$^{+	3 	}	_{-	2 	}$	&	0.80 	$^{+	0.04 	}	_{-	0.03 	}$	&	2.20 	$^{+	0.36 	}	_{-	0.35 	}$	&	100 	$^{+	5 	}	_{-	4 	}$	&	3.0 	$^{+	0.2 	}	_{-	0.2 	}$	&	5.9 	$^{+	0.3 	}	_{-	0.2 	}$	&	3.2E-18	\\
 BL Lacertae	&	H	&	0.0686	&	6.31E+15	&	0.11 	$^{+	0.04 	}	_{-	0.02 	}$	&	24 	$^{+	1 	}	_{-	3 	}$	&	0.91 	$^{+	0.10 	}	_{-	0.10 	}$	&	2.12 	$^{+	0.13 	}	_{-	0.21 	}$	&	50 	$^{+	3 	}	_{-	3 	}$	&	3.6 	$^{+	0.2 	}	_{-	0.2 	}$	&	5.9 	$^{+	0.4 	}	_{-	0.6 	}$	&	1.9E-16	\\
	&	L	&		&	9.77E+15	&	0.32 	$^{+	0.16 	}	_{-	0.17 	}$	&	24 	$^{+	4 	}	_{-	5 	}$	&	0.76 	$^{+	0.14 	}	_{-	0.13 	}$	&	1.98 	$^{+	0.22 	}	_{-	0.17 	}$	&	85 	$^{+	4 	}	_{-	4 	}$	&	2.9 	$^{+	0.3 	}	_{-	0.3 	}$	&	5.9 	$^{+	0.2 	}	_{-	0.3 	}$	&	1.4E-17	\\
 1ES 1011+496 	&	H	&	0.212	&	3.24E+15	&	0.96 	$^{+	0.11 	}	_{-	0.12 	}$	&	28 	$^{+	2 	}	_{-	2 	}$	&	0.85 	$^{+	0.03 	}	_{-	0.03 	}$	&	2.30 	$^{+	0.26 	}	_{-	0.24 	}$	&	15 	$^{+	3 	}	_{-	4 	}$	&	4.1 	$^{+	0.2 	}	_{-	0.2 	}$	&	6.6 	$^{+	0.5 	}	_{-	0.1 	}$	&	2.6E-15	\\
	&	L	&		&	5.01E+15	&	0.89 	$^{+	0.09 	}	_{-	0.10 	}$	&	17 	$^{+	4 	}	_{-	5 	}$	&	0.78 	$^{+	0.03 	}	_{-	0.06 	}$	&	1.87 	$^{+	0.30 	}	_{-	0.26 	}$	&	44 	$^{+	2 	}	_{-	2 	}$	&	3.5 	$^{+	0.1 	}	_{-	0.2 	}$	&	6.1 	$^{+	0.4 	}	_{-	0.6 	}$	&	2.4E-16	\\
 MAGIC J2001+435	&	H	&	0.18	&	5.25E+15	&	0.18 	$^{+	0.06 	}	_{-	0.03 	}$	&	25 	$^{+	3 	}	_{-	2 	}$	&	0.98 	$^{+	0.12 	}	_{-	0.20 	}$	&	2.05 	$^{+	0.37 	}	_{-	0.31 	}$	&	75 	$^{+	3 	}	_{-	5 	}$	&	3.7 	$^{+	0.2 	}	_{-	0.2 	}$	&	5.9 	$^{+	0.5 	}	_{-	0.7 	}$	&	2.9E-16	\\
	&	L	&		&	1.26E+16	&	0.14 	$^{+	0.04 	}	_{-	0.02 	}$	&	27 	$^{+	2 	}	_{-	2 	}$	&	1.01 	$^{+	0.05 	}	_{-	0.06 	}$	&	2.20 	$^{+	0.19 	}	_{-	0.20 	}$	&	45 	$^{+	4 	}	_{-	5 	}$	&	3.4 	$^{+	0.2 	}	_{-	0.2 	}$	&	6.6 	$^{+	0.2 	}	_{-	0.3 	}$	&	1.0E-17	\\
Mkn180	&		&	0.045278	&	1.29E+15	&	0.33 	$^{+	0.05 	}	_{-	0.04 	}$	&	22 	$^{+	4 	}	_{-	4 	}$	&	0.65 	$^{+	0.16 	}	_{-	0.13 	}$	&	1.63 	$^{+	0.34 	}	_{-	0.32 	}$	&	50 	$^{+	3 	}	_{-	4 	}$	&	3.5 	$^{+	0.2 	}	_{-	0.3 	}$	&	6.1 	$^{+	0.8 	}	_{-	0.7 	}$	&		\\
 3C 66A	&		&	0.444	&	1.23E+16	&	0.16 	$^{+	0.08 	}	_{-	0.05 	}$	&	30 	$^{+	4 	}	_{-	3 	}$	&	0.90 	$^{+	0.10 	}	_{-	0.11 	}$	&	2.05 	$^{+	0.18 	}	_{-	0.17 	}$	&	100 	$^{+	5 	}	_{-	4 	}$	&	3.5 	$^{+	0.3 	}	_{-	0.2 	}$	&	6.1 	$^{+	0.4 	}	_{-	0.3 	}$	&		\\
PKS 1424+240	&		&	0.16	&	5.01E+15	&	0.38 	$^{+	0.18 	}	_{-	0.11 	}$	&	27 	$^{+	4 	}	_{-	5 	}$	&	1.10 	$^{+	0.14 	}	_{-	0.16 	}$	&	2.70 	$^{+	0.28 	}	_{-	0.12 	}$	&	45 	$^{+	5 	}	_{-	4 	}$	&	3.8 	$^{+	0.2 	}	_{-	0.2 	}$	&	5.9 	$^{+	0.5 	}	_{-	0.3 	}$	&		\\
1ES 1215+303	&		&	0.13	&	1.38E+16	&	0.09 	$^{+	0.03 	}	_{-	0.02 	}$	&	27 	$^{+	4 	}	_{-	3 	}$	&	0.53 	$^{+	0.12 	}	_{-	0.13 	}$	&	1.35 	$^{+	0.19 	}	_{-	0.16 	}$	&	90 	$^{+	2 	}	_{-	2 	}$	&	2.5 	$^{+	0.3 	}	_{-	0.3 	}$	&	6.3 	$^{+	0.3 	}	_{-	0.4 	}$	&		\\
1RXS J003334.6-192130	&		&	0.61	&	1.00E+16	&	0.65 	$^{+	0.13 	}	_{-	0.18 	}$	&	21 	$^{+	3 	}	_{-	4 	}$	&	0.55 	$^{+	0.14 	}	_{-	0.11 	}$	&	1.40 	$^{+	0.20 	}	_{-	0.20 	}$	&	85 	$^{+	3 	}	_{-	3 	}$	&	2.7 	$^{+	0.2 	}	_{-	0.3 	}$	&	6.6 	$^{+	0.6 	}	_{-	0.8 	}$	&		\\
RGB J0152+017	&		&	0.08	&	2.51E+15	&	0.20 	$^{+	0.15 	}	_{-	0.10 	}$	&	20 	$^{+	3 	}	_{-	3 	}$	&	0.36 	$^{+	0.12 	}	_{-	0.19 	}$	&	1.70 	$^{+	0.26 	}	_{-	0.22 	}$	&	80 	$^{+	4 	}	_{-	3 	}$	&	3.2 	$^{+	0.2 	}	_{-	0.3 	}$	&	6.6 	$^{+	0.5 	}	_{-	0.4 	}$	&		\\
 1ES 0414+009 	&		&	0.287	&	7.76E+15	&	0.62 	$^{+	0.09 	}	_{-	0.09 	}$	&	23 	$^{+	5 	}	_{-	4 	}$	&	0.59 	$^{+	0.09 	}	_{-	0.09 	}$	&	1.72 	$^{+	0.31 	}	_{-	0.30 	}$	&	10 	$^{+	1 	}	_{-	4 	}$	&	3.3 	$^{+	0.2 	}	_{-	0.1 	}$	&	6.6 	$^{+	0.3 	}	_{-	0.5 	}$	&		\\
 PKS 0447-439	&		&	0.107	&	7.59E+15	&	0.52 	$^{+	0.13 	}	_{-	0.15 	}$	&	14 	$^{+	3 	}	_{-	3 	}$	&	0.82 	$^{+	0.16 	}	_{-	0.14 	}$	&	1.51 	$^{+	0.25 	}	_{-	0.26 	}$	&	30 	$^{+	6 	}	_{-	4 	}$	&	3.3 	$^{+	0.4 	}	_{-	0.2 	}$	&	5.9 	$^{+	0.3 	}	_{-	0.2 	}$	&		\\
RGB J0710+591	&		&	0.125	&	1.58E+15	&	0.72 	$^{+	0.14 	}	_{-	0.13 	}$	&	28 	$^{+	4 	}	_{-	3 	}$	&	0.62 	$^{+	0.04 	}	_{-	0.04 	}$	&	2.40 	$^{+	0.20 	}	_{-	0.22 	}$	&	20 	$^{+	3 	}	_{-	2 	}$	&	4.3 	$^{+	0.4 	}	_{-	0.2 	}$	&	6.0 	$^{+	0.3 	}	_{-	0.1 	}$	&		\\
1ES 0806+524 	&		&	0.138	&	2.82E+15	&	0.57 	$^{+	0.10 	}	_{-	0.11 	}$	&	20 	$^{+	2 	}	_{-	3 	}$	&	0.90 	$^{+	0.11 	}	_{-	0.14 	}$	&	1.68 	$^{+	0.15 	}	_{-	0.20 	}$	&	37 	$^{+	4 	}	_{-	3 	}$	&	3.5 	$^{+	0.2 	}	_{-	0.2 	}$	&	6.4 	$^{+	0.2 	}	_{-	0.2 	}$	&		\\
 MS 1221.8+2452	&		&	0.183648	&	6.31E+15	&	0.29 	$^{+	0.16 	}	_{-	0.10 	}$	&	20 	$^{+	3 	}	_{-	1 	}$	&	0.52 	$^{+	0.10 	}	_{-	0.10 	}$	&	1.78 	$^{+	0.19 	}	_{-	0.17 	}$	&	60 	$^{+	2 	}	_{-	3 	}$	&	3.4 	$^{+	0.2 	}	_{-	0.1 	}$	&	6.5 	$^{+	0.4 	}	_{-	0.2 	}$	&		\\
H 1426+428	&		&	0.129139	&	2.51E+15	&	0.11 	$^{+	0.08 	}	_{-	0.04 	}$	&	32 	$^{+	3 	}	_{-	3 	}$	&	0.36 	$^{+	0.11 	}	_{-	0.10 	}$	&	1.20 	$^{+	0.30 	}	_{-	0.33 	}$	&	90 	$^{+	5 	}	_{-	4 	}$	&	2.8 	$^{+	0.4 	}	_{-	0.4 	}$	&	6.0 	$^{+	0.4 	}	_{-	0.3 	}$	&		\\
PG 1553+113	&		&	0.36	&	3.89E+15	&	0.96 	$^{+	0.18 	}	_{-	0.19 	}$	&	27 	$^{+	2 	}	_{-	4 	}$	&	0.98 	$^{+	0.10 	}	_{-	0.09 	}$	&	2.61 	$^{+	0.19 	}	_{-	0.18 	}$	&	13 	$^{+	3 	}	_{-	2 	}$	&	4.3 	$^{+	0.4 	}	_{-	0.3 	}$	&	6.7 	$^{+	0.5 	}	_{-	0.4 	}$	&		\\
 B3 2247+381 	&		&	0.1187	&	3.98E+15	&	0.75 	$^{+	0.15 	}	_{-	0.19 	}$	&	17 	$^{+	4 	}	_{-	3 	}$	&	0.84 	$^{+	0.10 	}	_{-	0.11 	}$	&	1.81 	$^{+	0.34 	}	_{-	0.34 	}$	&	5 	$^{+	2 	}	_{-	2 	}$	&	3.7 	$^{+	0.1 	}	_{-	0.2 	}$	&	6.7 	$^{+	0.6 	}	_{-	0.5 	}$	&		\\
1ES 1741+196	&		&	0.084	&	3.98E+16	&	0.65 	$^{+	0.15 	}	_{-	0.15 	}$	&	8 	$^{+	4 	}	_{-	4 	}$	&	0.31 	$^{+	0.06 	}	_{-	0.05 	}$	&	1.60 	$^{+	0.13 	}	_{-	0.16 	}$	&	23 	$^{+	2 	}	_{-	2 	}$	&	2.2 	$^{+	0.2 	}	_{-	0.1 	}$	&	6.1 	$^{+	0.5 	}	_{-	0.5 	}$	&		\\
1ES 1218+30.4	&		&	0.183648	&	1.45E+15	&	0.65 	$^{+	0.12 	}	_{-	0.18 	}$	&	31 	$^{+	3 	}	_{-	5 	}$	&	1.06 	$^{+	0.13 	}	_{-	0.12 	}$	&	2.25 	$^{+	0.19 	}	_{-	0.17 	}$	&	52 	$^{+	5 	}	_{-	2 	}$	&	4.0 	$^{+	0.2 	}	_{-	0.2 	}$	&	6.0 	$^{+	0.3 	}	_{-	0.6 	}$	&		\\
H 2356-309	&		&	0.165388	&	2.29E+15	&	0.59 	$^{+	0.13 	}	_{-	0.17 	}$	&	18 	$^{+	2 	}	_{-	4 	}$	&	0.57 	$^{+	0.10 	}	_{-	0.11 	}$	&	1.42 	$^{+	0.14 	}	_{-	0.11 	}$	&	49 	$^{+	4 	}	_{-	4 	}$	&	3.3 	$^{+	0.2 	}	_{-	0.2 	}$	&	6.2 	$^{+	0.5 	}	_{-	0.4 	}$	&		\\
\hline
\end{tabular}
\end{minipage}
\flushleft{\textbf{Notes.} Column (12) is the observed or extrapolated flux density at 1 TeV.}
\end{table*}

\begin{table*}
\centering
\begin{minipage}{250mm}
\caption{The jet powers in the different forms ($P_{\text{e}}$, $P_{\text{p}}$, $P_{\text{B}}$), the radiative powers ($P_{\text{r}}$), and the black hold masses}
\scriptsize
\begin{tabular}{@{}ccccccccc@{}}
\hline\hline
Name & State & $P_{\text{e}}$~(erg~s$^{-1}$) & $P_{\text{p}}$~(erg~s$^{-1}$) & $P_{\text{B}}$~(erg~s$^{-1}$) & $P_{\text{r}}$~(erg~s$^{-1}$) & $\log(M_{BH}/M_{\odot})$ & Ref \\
{(1)} & {(2)} & {(3)} & {(4)} & {(5)} & {(6)} & {(7)} & {(8)} \\
\hline
1ES 1101-232	&	H	&	9.88E+43	&	3.03E+44	&	4.09E+42	&	3.89E+42	&		&		\\
	&	L	&	9.55E+42	&	3.75E+42	&	1.87E+43	&	1.06E+43	&		&		\\
Mkn421	&	H	&	3.48E+43	&	1.15E+43	&	2.13E+42	&	2.26E+43	&	8.5, 8.29, 8.22, 7.6	&	Sb12, W02, C12, X04	\\
	&	L	&	1.17E+43	&	3.78E+42	&	5.11E+42	&	1.65E+42	&		&		\\
Mkn501	&	H	&	1.77E+43	&	2.35E+42	&	1.42E+41	&	8.30E+42	&	9, 9.21, 8.3	&	Sb12, W02, X04	\\
	&	L	&	1.18E+43	&	9.41E+42	&	5.76E+41	&	8.34E+41	&		&		\\
1ES 1959+650	&	H	&	3.27E+43	&	6.83E+42	&	8.27E+41	&	7.66E+42	&	8.1	&	Z12	\\
	&	L	&	5.65E+42	&	2.17E+42	&	8.66E+42	&	6.85E+42	&		&		\\
PKS 2005-489	&	H	&	1.09E+44	&	9.01E+44	&	3.40E+43	&	3.40E+42	&	8.1	&	X04	\\
	&	L	&	1.50E+44	&	1.20E+44	&	7.34E+41	&	1.79E+42	&		&		\\
PKS 2155-304	&	H	&	3.16E+44	&	1.13E+44	&	5.59E+42	&	2.91E+44	&	8.7	&	Z12	\\
	&	L	&	3.45E+43	&	2.15E+43	&	9.17E+42	&	1.86E+43	&		&		\\
1ES 2344+514	&	H	&	1.08E+43	&	1.73E+42	&	1.33E+41	&	2.54E+42	&	8.8	&	W02	\\
	&	L	&	1.18E+43	&	6.39E+42	&	2.28E+41	&	3.42E+41	&		&		\\
 W Comae	&	H	&	2.70E+44	&	2.32E+44	&	9.26E+41	&	6.51E+42	&	8	&	X04	\\
	&	L	&	6.89E+43	&	7.87E+43	&	4.25E+42	&	9.07E+42	&		&		\\
S5 0716+714	&	H	&	7.31E+44	&	7.39E+44	&	5.17E+43	&	2.08E+44	&	8.1, 8.1	&	C12, X04	\\
	&	L	&	3.67E+44	&	1.83E+45	&	4.18E+44	&	2.15E+44	&		&		\\
 BL Lacertae	&	H	&	1.65E+44	&	1.66E+44	&	1.07E+42	&	3.09E+42	&	8.7, 8.23, 7.7	&	Sb12, W02, X04	\\
	&	L	&	1.43E+44	&	6.79E+44	&	2.16E+43	&	4.58E+42	&		&		\\
 1ES 1011+496 	&	H	&	3.90E+43	&	1.68E+43	&	2.73E+43	&	4.26E+43	&	8.3	&	Z12	\\
	&	L	&	4.95E+43	&	4.63E+43	&	2.23E+43	&	4.86E+43	&		&		\\
 MAGIC J2001+435	&	H	&	2.52E+44	&	1.84E+44	&	1.99E+42	&	2.02E+43	&		&		\\
	&	L	&	4.89E+44	&	7.58E+44	&	8.10E+42	&	1.33E+43	&		&		\\
Mkn180	&		&	1.05E+44	&	1.36E+44	&	5.01E+42	&	2.53E+43	&	8.21	&	W02	\\
 3C 66A	&		&	1.76E+45	&	1.93E+45	&	1.31E+43	&	3.16E+44	&	8.6, 8	&	C12, X04	\\
PKS 1424+240	&		&	1.20E+44	&	1.17E+44	&	9.91E+42	&	1.33E+43	&		&		\\
1ES 1215+303	&		&	4.40E+44	&	1.78E+45	&	4.22E+42	&	5.31E+42	&	8.12	&	W02	\\
1RXS J003334.6-192130	&		&	1.76E+44	&	5.14E+44	&	6.65E+43	&	1.01E+44	&		&		\\
RGB J0152+017	&		&	1.40E+43	&	2.90E+43	&	3.78E+41	&	5.79E+41	&		&		\\
 1ES 0414+009 	&		&	3.33E+43	&	4.59E+43	&	4.65E+43	&	1.64E+43	&		&		\\
 PKS 0447-439	&		&	4.87E+43	&	4.87E+43	&	1.16E+43	&	2.21E+43	&		&		\\
RGB J0710+591	&		&	1.00E+43	&	5.50E+42	&	3.77E+42	&	3.12E+42	&	8.26	&	W02	\\
1ES 0806+524 	&		&	2.18E+43	&	1.59E+43	&	3.72E+42	&	4.47E+42	&	8.9	&	Z12	\\
 MS 1221.8+2452	&		&	7.71E+42	&	5.77E+42	&	3.13E+41	&	3.17E+41	&		&		\\
H 1426+428	&		&	7.55E+43	&	8.24E+43	&	2.93E+41	&	2.46E+42	&	9.13	&	W02	\\
PG 1553+113	&		&	5.72E+43	&	2.03E+43	&	3.81E+43	&	8.57E+43	&		&		\\
 B3 2247+381 	&		&	6.15E+42	&	3.29E+42	&	9.79E+42	&	3.18E+42	&		&		\\
1ES 1741+196	&		&	3.46E+43	&	3.47E+44	&	1.68E+44	&	4.54E+43	&		&		\\
1ES 1218+30.4	&		&	3.59E+43	&	1.49E+43	&	3.20E+42	&	8.61E+42	&	8.58	&	W02	\\
H 2356-309	&		&	1.51E+43	&	1.18E+43	&	2.22E+42	&	4.99E+42	&	8.6	&	W02	\\
\hline
\end{tabular}
\end{minipage}
\flushleft{\textbf{Notes.} Column (8) is the references of black hole masses. C12: Chai et al. (2012); Sb12: Shaw et al. (2012); W02: Woo \& Urry (2002); X04: Xie et al. (1991, 2004); Z12: Zhang et al. (2012).}
\end{table*}

\begin{table*}
\centering
\begin{minipage}{250mm}
\caption{The results of covariance analysis and partial correlation analysis for $B$-$\delta$ correlation and $r$-$s$ correlation}
\scriptsize
\begin{tabular}{|c|c|cccccc}
\hline\hline
\multirow{4}{*}{$B$-$\delta$ correlation}
& covariate$^{\text{a}}$	&	$R$	&	$N$	&	$r$	&	$s$	&	$\gamma_{0}$	&	$\gamma_{\max}$	\\
\cline{2-2}
& Chance probability of covariance analysis($P_{\text{cov}}$)$^{\text{b}}$	&	0.06	&	0.53	&	0.10	&	0.35	&	0.08	&	0.76	\\
\cline{2-2}
& Pearson coefficient of partial correlation analysis ($\rho_{\text{par}}$)$^{\text{c}}$	&	-0.48	&	-0.42	&	-0.49	&	-0.52	&	-0.47	&	-0.48	\\
\cline{2-2}
& Chance probability of partial correlation analysis ($P_{\text{par}}$)$^{\text{d}}$	&	0.002	&	0.006	&	0.001	&	0.001	&	0.002	&	0.001	\\
\hline
\multirow{4}{*}{$r$-$s$ correlation}
& covariate	&	$B$	&	$\delta$	&	$R$	&	$N$	&	$\gamma_{0}$	&	$\gamma_{\max}$	\\
\cline{2-2}
& Chance probability of covariance analysis ($P_{\text{cov}}$)	&	0.19	&	0.29	&	0.18	&	0.43	&	0.42	&	0.42	\\
\cline{2-2}
& Pearson coefficient of partial correlation analysis ($\rho_{\text{par}}$)	&	0.73	&	0.69	&	0.70	&	0.70	&	0.59	&	0.69	\\
\cline{2-2}
& Chance probability of partial correlation analysis ($P_{\text{par}}$)	&	$<0.001$	&	$<0.001$	&	$<0.001$	&	$<0.001$	&	$<0.001$	&	$<0.001$	\\
\hline
\end{tabular}
\end{minipage}
\flushleft{\textbf{Notes.} a. Possible covariate in correlation analysis ($B$-$\delta$ correlation and $r$-$s$ correlation). b. The results of covariance analysis. Null hypothesis: there exists the influence of covariate (i.e., If the chance probability of covariance analysis $P_{\text{cov}}$ < 0.05, there exists the influence of covariate). c. The Pearson correlation coefficient excluding the influence of covariate. d. The chance probability of correlation excluding the influence of covariate. If the chance probability of partial correlation analysis $P_{\text{par}}$ < 0.05, there exists the correlation between dependent variable and independent variable excluding the influence of covariate (Machalski \& Jamrozy 2006).}
\end{table*}
\end{document}